\documentclass[%
 reprint,
nofootinbib,
 amsmath,amssymb,
 aps,
showkeys
]{revtex4-2}

\usepackage{graphicx}
\usepackage{dcolumn}
\usepackage{bm}
\usepackage{amsmath}
\usepackage{amssymb}
\usepackage{braket}
\usepackage[version=3]{mhchem}
\usepackage{xfrac}
\usepackage{multirow}
\usepackage{xcolor}

\begin{document}

\preprint{APS/123-QED}

\title{Comparative study of magnetocaloric properties for Gd$^{3+}$ compounds with different frustrated lattice geometries}

\author{EliseAnne C. Koskelo} 
\altaffiliation[Now at ]{Department of Physics, Harvard University, Cambridge, MA 02138, USA.}
\email{eck34@cantab.ac.uk}
\author{Paromita Mukherjee} 
\author{Cheng Liu}
\author{Alice C. Sackville Hamilton}
\author{Harapan S. Ong}
\author{Claudio Castelnovo}
\author{Si\^{a}n E. Dutton}
\email{sed33@cam.ac.uk}
\affiliation{
Department of Physics, University of Cambridge, Cambridge CB3 0HE, United Kingdom }

\author{M. E. Zhitomirsky}
\affiliation{
 Universit\'{e} Grenoble Alpes, CEA, IRIG, PHELIQS, 38000 Grenoble, France
}

\date{\today}

\begin{abstract}
As materials with suppressed ordering temperatures and enhanced ground state entropies, frustrated magnetic oxides are ideal candidates for cryogenic magnetocaloric refrigeration. While previous materials design has focused on tuning the magnetic moments, their interactions, and density of moments on the lattice, there has been relatively little attention to frustrated lattices. Prior theoretical work has shown that the magnetocaloric cooling rate at the saturation field is proportional to a macroscopic number of soft mode excitations that arise due to the classical ground state degeneracy. The number of these modes is directly determined by the geometry of the frustrating lattice. For corner-sharing geometries, the pyrochlore has 50\% more modes than the garnet and kagome lattices, whereas the edge-sharing \emph{fcc} has only a subextensive number of soft modes. Here, we study the role of soft modes in the magnetocaloric effect of four large-spin Gd$^{3+}$ ($L=0$, $J=S=7/2$) Heisenberg antiferromagnets on a kagome, garnet, pyrochlore, and \emph{fcc} lattice. By comparing measurements of the magnetic entropy change $\Delta S_m$ of these materials at fields up to $9$~T with predictions using mean-field theory and Monte Carlo simulations, we are able to understand the relative importance of spin correlations and quantization effects. We observe that tuning the value of the nearest neighbor coupling has a more dominant contribution to the magnetocaloric entropy change in the liquid-He cooling regime ($2$-$20$~K), rather than tuning the number of soft mode excitations. Our results inform future materials design in terms of dimensionality, degree of magnetic frustration, and lattice geometry.
\end{abstract}

\keywords{magnetocaloric effect, magnetic frustration, lanthanide oxides, soft modes, superexchange}
\maketitle

\section{Introduction}

The ever-increasing need for cooling in modern technologies, as well as the scarcity of helium, has motivated the search for sustainable cooling alternatives, including magnetocaloric materials~\cite{Science_Moya_Caloric}. The magnetocaloric effect describes the adiabatic temperature change of a magnetic material due to a change in applied field. Common early refrigerants were based on dilute paramagnetic salts such as cerous magnesium nitrate (CMN) and ferric ammonium alum (FAA)~\cite{MCE_review_PMsalts}. Research interests shifted later on from dilute paramagnetic salts and superparamagnets to dense magnetic lattices magnets, when it was shown that frustration can enable an enhanced magnetocaloric effect via: 1) a suppressed ordering temperature at the same magnetic density and 2) a large ground state entropy/soft magnon modes~\cite{Zhitomirsky_2003}. Perhaps the most well-known frustrated magnetocaloric material, Gd$_3$Ga$_5$O$_{12}$, exhibits a magnetic entropy difference of $13.0$~J K$^{-1}$ mol$_{\textrm{Gd}}^{-1}$ in a field change from $7$~T to zero field at $2$~K, with cooling capabilities down to $\sim 0.8$~K~\cite{GdF3_MCE_2015,GGG_limits_MCE}. Other notable Gd$^{3+}$ examples include the inorganic frameworks GdF$_3$ ($15.3$~J K$^{-1}$ mol$_{\textrm{Gd}}^{-1}$) and Gd(OH)F$_2$ ($16.3$~J K$^{-1}$ mol$_{\textrm{Gd}}^{-1}$), the dense metal organic framework Gd(HCOO)$_3$ ($16.3$~J K$^{-1}$ mol$_{\textrm{Gd}}^{-1}$), and the frustrated monazite antiferromagnet GdPO$_4$ (15.6 J K$^{-1}$ mol$_{\textrm{Gd}}^{-1}$), all again in a field change from $7$~T to zero field at $2$~K ~\cite{GdF3_MCE_2015, xu_gdohf2_2022,Gd-formate_2013, GdPO4_MCE_2014}.

In the case of Heisenberg Gd$^{3+}$, current magnetocaloric research for cooling in the liquid He regime has focused on minimizing anisotropy of magnetic ions and their superexchange interactions to maximize spin polarizability~\cite{GdF3_MCE_2015, MCE_review_2020,Gd-formate_2013}. However, using the subtle effects of magnetic frustration to enhance the magnetocaloric effect and to suppress the magnetic ordering temperature remains a relatively unexplored area of research~\cite{Paromita_thesis, YbGG_MCE}. In frustrated magnetic oxides, the main tuning parameters are the magnetic ions and the atomic lattice. In this way, the superexchange and dipolar interactions, and the crystal electric field (CEF) environment of individual ions are manipulated.

\begin{figure}[t]
    \centering
    \includegraphics[width=0.9\columnwidth]{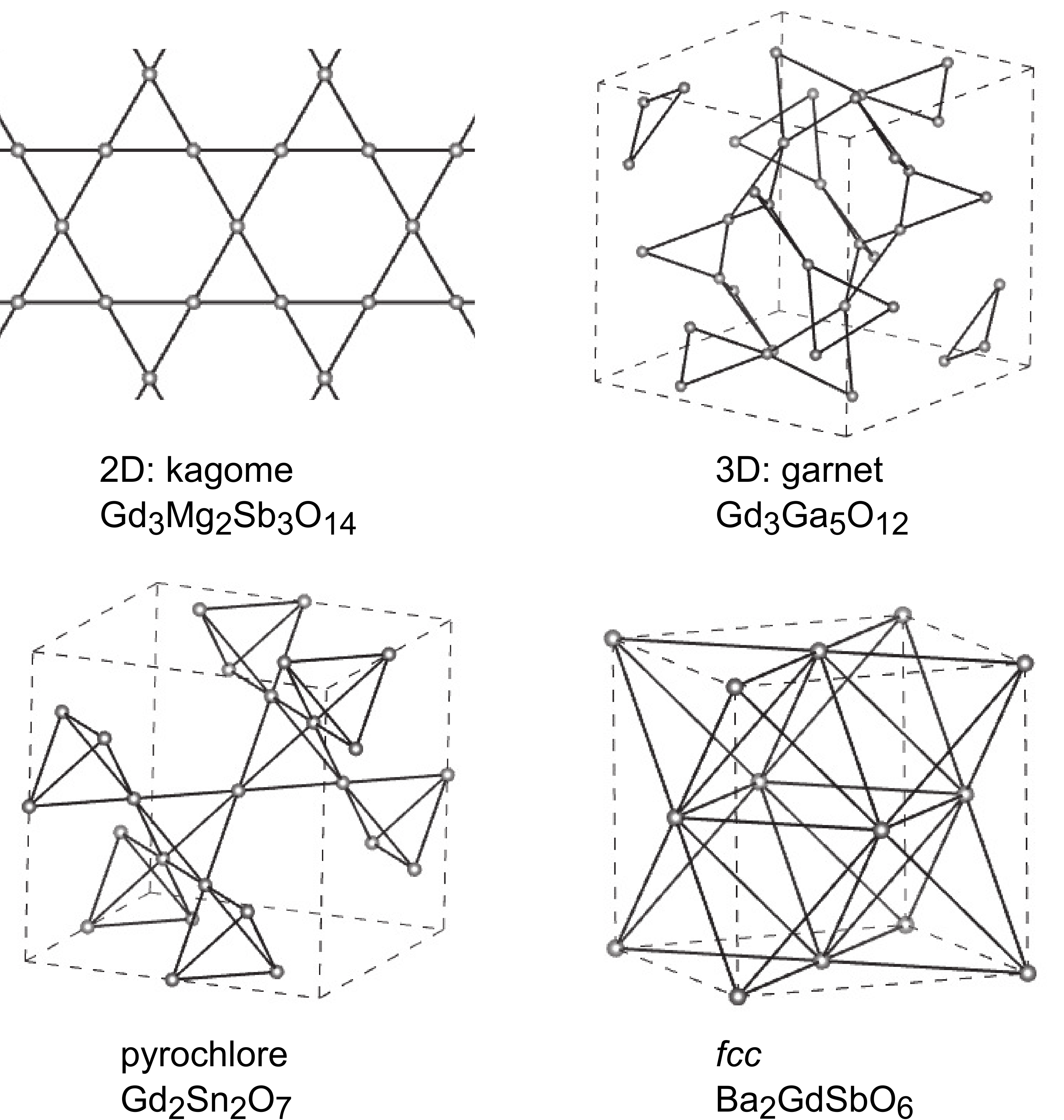}
    \caption{Frustrated lattice geometries in 2D (kagome) and 3D (garnet, pyrochlore, \emph{fcc}).}
    \label{fig:geometries}
\end{figure}

In the case where dipolar and CEF contributions are negligible, prior theoretical work has shown that the lattice geometry can dictate an enhancement of the magnetocaloric cooling rate, $(\partial T/\partial H)_{S_m} \propto -(\partial S_m/\partial H)_T$, that scales with the macroscopic number of soft modes $N_4$ at the saturation field~\cite{Zhitomirsky_2003}. The soft modes are ultimately a result of the macroscopic ground state entropy that frustration introduces. These modes have been modeled for the three corner-sharing geometries shown in Figure~\ref{fig:geometries}: the pyrochlore lattice, with the number of pyrochlore modes scaling as $N$, and the garnet and kagome lattices with $2N/3$ soft modes, where $N$ is the number of lattice sites~\cite{Zhitomirsky_2003}. Figure~\ref{fig:geometries} also shows an edge-sharing geometry: the \emph{fcc} lattice, for which the number of soft modes has not been reported.

In this work, we seek to test experimentally the role of these fundamental soft modes in the measured magnetocaloric effect in four representative frustrated oxide materials in the liquid He regime ($2$-$20$~K). Of particular interest is answering the question of the optimal frustrating geometry to inform future magnetocaloric material design. Four Gd$^{3+}$-based oxides were chosen as model systems since the contribution of crystal electric field effects to the magnetism are negligible ($L=0$ and hence $J=S=7/2$ and $g_J=g=2$). The four compounds investigated include Gd$_3$Mg$_2$Sb$_3$O$_{14}$, in which the Gd$^{3+}$ ions lie on kagome layers separated by triangular layers of nonmagnetic Mg$^{2+}$ ions~\cite{Dun_LnMSO_2016_PRL}; Gd$_3$Ga$_5$O$_{12}$, in which Gd$^{3+}$ ions lie on two interpenetrating rings of triangles~\cite{Petrenko_GGG_Dipolar_term_2000}; Gd$_2$Sn$_2$O$_7$, in which the Gd$^{3+}$ ions form a pyrochlore lattice (corner-sharing tetrahedra); and Ba$_2$GdSbO$_6$, an \emph{fcc} lattice of Gd$^{3+}$ ions (edge-sharing tetrahedra). 

We find that, upon normalizing for the differences in superexchange across the materials, the normalized magnetic entropy change $\sqrt{J_1} \Delta S_m$ in this temperature range qualitatively scales with the number of soft modes for the three-dimensional lattices as predicted in Ref.~\onlinecite{Zhitomirsky_2003}. However, we find that the \emph{fcc} lattice exhibits a 30-95\% greater magnetic entropy change $\Delta S_m$ than the corner-sharing lattices, which can be attributed to its paramagnetic response~\cite{Koskelo_2022}. 
We compare the experimental results with Monte Carlo simulations of classical Heisenberg spins as well as with mean-field superexchange calculations that account for the quantised nature of the spins. 
These allow us to show that, in the temperature and field regimes of interest in this study, the effect of correlations is dominant over quantization in corner sharing compounds, whereas the opposite is true for the \emph{fcc} case. 
The fact that the magnetocaloric cooling rate of the corner-sharing geometries cannot be fully described using a mean-field superexchange model is consistent with a significant contribution due to soft modes to the total magnetic entropy change measured. 

For the compounds studied, we find that the paramagnetic contribution to the magnetic entropy outweighs that of geometric frustration. In particular, the three-fold reduction in the superexchange $J_1$ of the garnet compound ($J_1 \sim 300$~mK~\cite{Koskelo_2022}) results in a 50\% improvement in the magnetic entropy change extracted compared to the pyrochlore lattice. These results suggest that compounds with the smallest superexchange $J_1$ and which also make use of magnetic frustration are promising future magnetocaloric materials. 
%
%

\section{Soft Modes Model: Validity in Liquid He Regime}

In prior work~\cite{Zhitomirsky_2003}, one of us predicted a strongly enhanced magnetocaloric 
effect for geometrically frustrated magnets with high classical degeneracy in the ground state. The work also included classical Monte Carlo simulations, generally valid for large-$S$ magnetic materials, for Heisenberg antiferromagnets with three typical frustrated geometries based on corner-sharing plaquettes: the two-dimensional kagome lattice (a network of corner-sharing triangles), the garnet lattice (its three-dimensional analog), and the pyrochlore lattice (a network of corner-sharing tetrahedra), see Figure~\ref{fig:geometries}. 
Here we revisit these geometries and compare them with an \emph{fcc} lattice. 

The spin Hamiltonian consists of a nearest neighbor ($nn$) superexchange term, assumed to be uniform across all $nn$ pairs, and a Zeeman term:
\begin{equation}
    \hat{\mathcal{H}} = J_1 \sum_{\braket{i j}} \mathbf{S}_i \cdot \mathbf{S}_j - \mathbf{H} \cdot \sum_i \mathbf{S}_i
    \, ,
    \label{eq:spinH}
\end{equation}
where $J_1$ is the $nn$ superexchange and a factor $g \mu_B$ has been absorbed into the definition of the field $H$ for convenience. 
Dipolar interactions are also present in the system but we shall disregard them in our simulations. Considering the system parameters reported later in Table~\ref{table:J1_D_Gd_soft_modes}, we expect this to be a reasonable approximation for the three corner sharing lattices. In the \emph{fcc} system, superexchange and dipolar interactions at nearest-neighbour distance are of similar strength; however, they are much smaller than the temperature regime of interest in our study and therefore truncating them to nearest-neighbour distance is a valid approximation. 

In the low-temperature cooperative paramagnetic state, a condensation of a macroscopic number $N_4$ of soft modes at the saturation field $H_{\rm sat}$ is predicted to occur, due to the enhanced ground state entropy afforded by the underconstraint of frustration~\cite{Zhitomirsky_2003}. The saturation field, defined as the threshold beyond which the energy becomes dominated by the Zeeman term, has been predicted~\cite{Zhitomirsky_2003} 
to be $H_{\rm sat} = 6 J_1 S$ for the kagome and garnet lattices, and $H_{\rm sat} = 8 J_1 S$ for the pyrochlore lattice, see Table~\ref{table:Hsat_Gd_compounds_soft_modes}. 
Using the same frustrated block decomposition~\cite{zhitomirsky2015real}, we obtain here the saturation field $H_{\rm sat} = 16 J_1 S$ for the nearest neighbor \emph{fcc} antiferromagnet (see Appendix~\ref{app:Hsat}).

\begin{table}[t]
    \caption{Predicted number of soft modes $N_4$ for a lattice with $N$ sites of Heisenberg spins and predicted saturation field $H_{\rm sat,theory}$~\cite{Zhitomirsky_2003} versus measured saturation field $H_{\rm sat,obs}$ for Gd$_3$Ga$_5$O$_{12}$, Gd$_3$Mg$_2$Sb$_3$O$_{14}$, Gd$_2$Sn$_2$O$_7$, and Ba$_2$GdSbO$_6$. The observed saturation field, $H_{\rm sat,obs}$, was determined as the field at which the measured temperature gradient of the magnetization $(\partial M/\partial T)_H$ is maximized at $T = 2$~K. We estimate the expected scaling for the number of soft modes for \emph{fcc} Ba$_2$GdSbO$_6$ based on the magnon spectrum at $H_{\rm sat}$ derived in Ref.~\onlinecite{fcc_magnons_atHsat_Zhitomirksy}. Error bars listed for the measured saturation field are based on the $0.2$~T step size for the $M(H)$ measurements.}
    \label{table:Hsat_Gd_compounds_soft_modes}
    \centering
    \begin{ruledtabular}
    \begin{tabular}{l c c c c}
    Compound & Lattice & $N_4$ & $H_{\rm sat,theory}$  & $H_{\rm sat,obs}$\\
     & & & (T) & (T) \\
    \hline \\
        \vspace{0.1cm}
        Gd$_2$Sn$_2$O$_7$ & Pyrochlore & $N$ & $\frac{8 J_1 S}{g \mu_B}$= 6.3 & 5.8(2) \\ \vspace{0.1cm}
        Gd$_3$Mg$_2$Sb$_3$O$_{14}$ & Kagome & $\frac{2}{3}N$ & $\frac{6 J_1 S}{g \mu_B}$ = 4.7 & 4.6(2) \\ \vspace{0.1cm}
        Gd$_3$Ga$_5$O$_{12}$ & Garnet & $\frac{2}{3}N$ & $\frac{6 J_1 S}{g \mu_B}$ = 1.7 & 2.0(2) \\ \vspace{0.1cm}
        Ba$_2$GdSbO$_6$ & \emph{fcc} & $\sim N^{1/3}$ & $\frac{16 J_1 S}{g \mu_B}$ = 0.5 & 1.4(2)\footnote{As we discuss in the text, the temperature in which the measurement is made may be too high for the value of $H_{\rm sat,obs}$ to be accurate in the \emph{fcc} Ba$_2$GdSbO$_6$ case.} \\
    \end{tabular}
    \end{ruledtabular}
\end{table}

The magnetocaloric rate, $(\partial S_m/\partial H)_T$, at the saturation field is predicted to follow the scaling relation:
\begin{equation}
   \left( \frac{\partial S_m}{\partial H} \right)_{T,N4} \propto -\frac{N_4}{\sqrt{J_1 T}}, \hspace{12pt} \textrm{at }H = H_{\textrm{sat}} \, .
   \label{eq:sing_Zhito}
\end{equation}
This is in addition to the conventional contribution to $(\partial S_m/\partial H)_T$ from the paramagnetic (non-frustrated) ordinary dispersive modes ($N_2$) of the system, 
\begin{equation}
     \left( \frac{\partial S_m}{\partial H} \right)_{T,N2} \propto -\frac{1}{2} \sum_{\mathbf{k}} \frac{1}{(H-H_{\rm sat})S+\epsilon(\mathbf{k})}\, ,
     \label{eq:N2_modes_Zhito}
\end{equation}
where the excitation energies $\epsilon(\mathbf{k})$ are nonnegative and vanish for $k$ corresponding to the propagation vectors of degenerate classical ground states in zero field~\cite{Zhitomirsky_2003}. 

For non-frustrated, three-dimensional Heisenberg magnets above the ordering temperature, $\left( \frac{\partial S_m}{\partial H} \right)_{T,N2}$ is the only contribution and there is no enhancement in the magnetocaloric cooling rate, which is a temperature-independent constant for $H = H_{\rm sat}$ ~\cite{Zhitomirsky_2003}. 
In frustrated systems where $N_4$ is macroscopic (i.e., $N_4 \propto N$, where $N$ is the number of lattice sites), one expects $\left( \frac{\partial S_m}{\partial H} \right)_{T,N4}$ to dominate at low enough temperatures (where it grows as $\sim 1/\sqrt{T}$). However, when $N_4 \propto N^{\alpha}$ with $0 < \alpha < 1$, as is the case for \emph{fcc} Ba$_2$GdSbO$_6$ (discussed below), there will always be a system size beyond which $\left( \frac{\partial S_m}{\partial H} \right)_{T,N2}$ dominates, for any given temperature: $N \gtrsim 1/T^{1/[2(1-\alpha)]}$.

Using classical spin-wave calculations in the saturated (collinear) state, the number of soft modes in the pyrochlore lattice was found to scale with the number of lattice sites, $N_4 \propto N$, Table~\ref{table:Hsat_Gd_compounds_soft_modes}~\cite{Zhitomirsky_2003}. 
The kagome and garnet lattices are predicted to have $2/3$ as many soft modes as the pyrochlore. Thus, it could be expected that the pyrochlore lattice may have a greater magnetocaloric effect than a comparable garnet or kagome lattice. Indeed prior work has shown that the pyrochlore Gd$_2$Ti$_2$O$_7$ exhibits an increased cooling rate compared to the garnet Gd$_3$Ga$_5$O$_{12}$, but the work does not address the overall magnetic entropy change (and hence total cooling) available from each material nor their differing superexchange~\cite{Gd2Ti2O7_MCE_Tad_Zhitomirksy}.

The number of soft modes for an \emph{fcc} lattice has not been reported, but is expected to be lower than that for the corner-sharing geometries, as its zero energy modes correspond to lines (rather than surfaces) in the Brillouin zone~\cite{zhitomirsky_highFields_2005}. The appropriate scaling for the number of \emph{fcc} soft modes can be predicted from the magnon spectrum at the saturation field reported in Ref.~\onlinecite{fcc_magnons_atHsat_Zhitomirksy}:
\begin{equation}
    \epsilon_{\mathbf{q}} = 2\left(1+\cos q_x\cos q_y+\cos q_x\cos q_z + \cos q_y \cos q_z \right) \, .
\end{equation}
The zeros of this equation correspond to soft modes and are given by $\mathbf{q} = (\pi,q,0)$ and the ``cubic related lines''~\cite{fcc_magnons_atHsat_Zhitomirksy}. Along a given dimension of a material with $N$ lattice sites, there are $N^{1/3}$ such $\mathbf{q}$ states for which this is the case. Thus, the \emph{fcc} lattice has a subextensive number of soft modes compared to the corner-sharing lattices. 

The soft modes theory described here is valid for classical (large-spin) Heisenberg magnets~\cite{Zhitomirsky_2003}. The lower limit of this temperature regime, $T^*$, is approximated by $T^* \approx J_1 S$, below which quantum statistics must be used to analyze spin excitations~\cite{fcc_orderbydisorder_2020}. Among the compounds studied, the superexchange is largest for the pyrochlore and kagome compounds, $J_1 \sim 0.3$~K in Table~\ref{table:J1_D_Gd_soft_modes}, corresponding to a limiting temperature of $T^* \sim 1$~K. Thus the liquid-He temperature range investigated here, $2-20$~K, should provide reasonable insight into the role of soft modes on the magnetocaloric effect, as classical statistics of magnon modes applies at these temperatures. 
%
%

\section{Experimental Results}
%
%

\subsection{Magnetic Characterization}

\begin{figure}[htbp]
    \centering
    \includegraphics[width=\columnwidth]{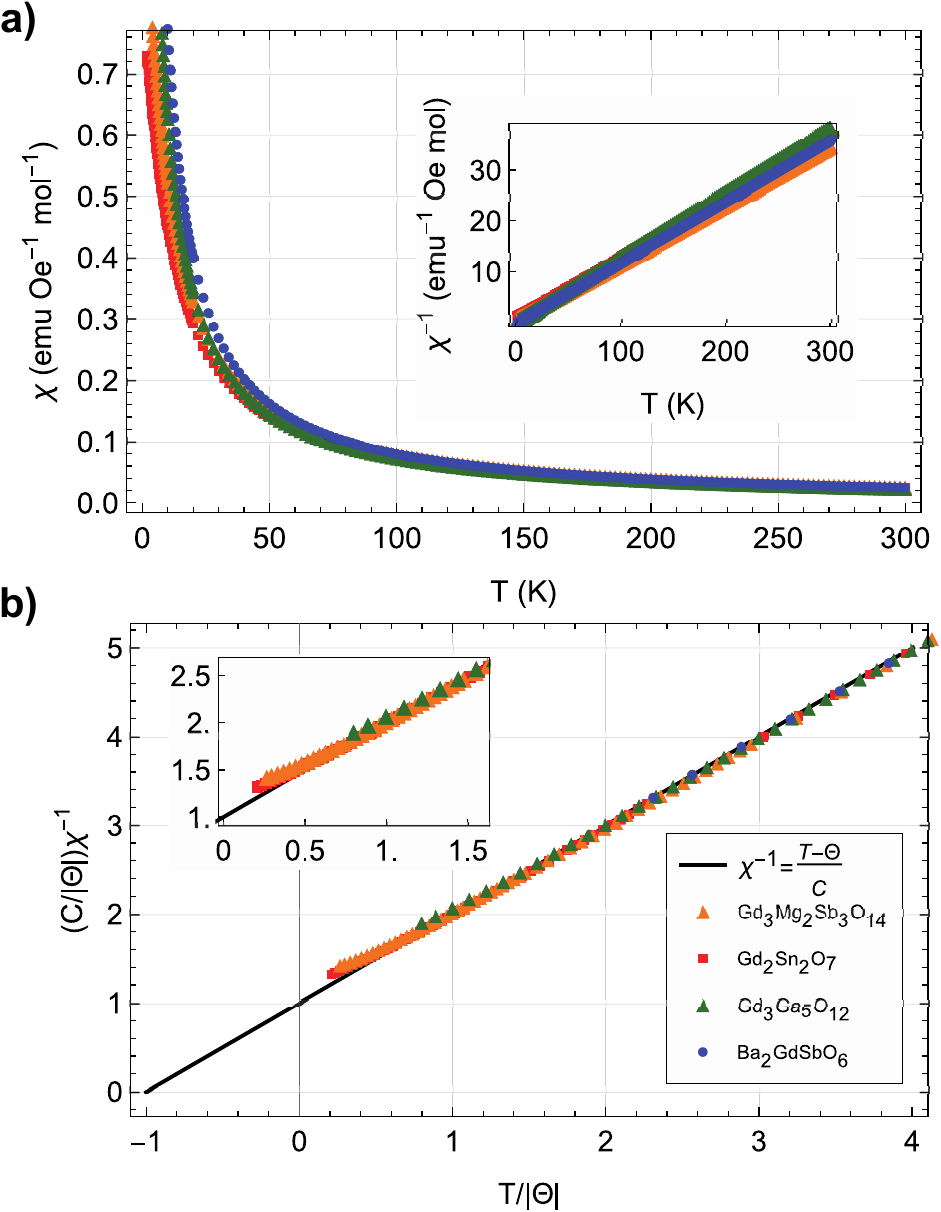}
    \caption{a) Measured static magnetic susceptibility $\chi$ of Gd-based kagome (Gd$_3$Mg$_2$Sb$_3$O$_{14}$), pyrochlore (Gd$_2$Sn$_2$O$_7$), garnet (Gd$_3$Ga$_5$O$_{12}$), and \emph{fcc} (Ba$_2$GdSbO$_6$) compounds versus temperature (inset: inverse magnetic susceptibility). b) Dimensionless inverse magnetic susceptibility versus dimensionless temperature (inset: zoomed-in region near zero temperature). Both axes are scaled using the appropriate factors of the Curie-Weiss temperature $\Theta$ and Curie constant $C$. All compounds exhibit positive deviations from paramagnetic behavior (black line) indicative of possible antiferromagnetic short-range correlations. The \emph{fcc} lattice is effectively paramagnetic with deviations of less than 5\% from Curie-Weiss behavior~\cite{Koskelo_2022}. All measurements were made over the same temperature range, $1.8-300$~K, but exhibit different dimensionless temperature ranges due to different $\Theta$.}
    \label{fig:inv_chi_Gdcompounds_soft_modes}
\end{figure}

The measured magnetic susceptibilities of Gd$_3$Mg$_2$Sb$_3$O$_{14}$, Gd$_2$Sn$_2$O$_7$, Gd$_3$Ga$_5$O$_{12}$, and Ba$_2$GdSbO$_6$ from $1.8$~K to $300$~K are depicted in Figure~\ref{fig:inv_chi_Gdcompounds_soft_modes}. Details of sample preparation and structural characterization can be found in Appendix~\ref{appendix:Gd_structures_soft_modes}. The negative Curie-Weiss temperatures, see Table~\ref{table:J1_D_Gd_soft_modes}, confirm the antiferromagnetic superexchange required for frustration and are consistent with previous literature reports~\cite{Dun_kagome_2017,Wellm_GMSO_2020,bondah-jagalu_magnetic_2001,Sackville_Hamilton_2014,bondah-jagalu_magnetic_2001}. Rearranging the Curie-Weiss law into the dimensionless form:
\begin{equation}
    \frac{C}{|\Theta| \chi}= \frac{T}{|\Theta|} + 1 \, , \hspace{10pt} \Theta<0 \, ,
\end{equation}
where $\Theta$ is the Curie temperature and $C$ is the Curie constant, we can compare the strength of magnetic short-range correlations between spins across different compounds~\cite{Melot_SRO_spinels_2009}. Positive deviations from the Curie-Weiss law indicate antiferromagnetic short-range correlations while negative deviations signify ferromagnetic correlations. As shown in Figure~\ref{fig:inv_chi_Gdcompounds_soft_modes}b), all corner-sharing geometries exhibit antiferromagnetic short-range correlations, while the \emph{fcc} compound is qualitatively paramagnetic down to $T = 1.8$~K, consistent with the literature~\cite{Karunadasa_PNAS,Paddison_Gd2Sn2O7_2017,Wellm_GMSO_2020,Koskelo_2022,Paddison_GGG_2015,Petrenko_GGG_Dipolar_term_2000}. A mean-field estimate for the $nn$ superexchange can be computed from the Curie-Weiss law using:
\begin{equation}
    J_1 = \frac{3|\Theta|}{z S(S+1)} \, ,
\end{equation} 
where $z$ is the number of $nn$. 

Table~\ref{table:J1_D_Gd_soft_modes} lists the reported values for the $nn$ superexchange $J_1$ and dipolar interaction $D$ for each compound which are in agreement with the values obtained from the Curie-Weiss fits of the measured magnetic susceptibility in Figure~\ref{fig:inv_chi_Gdcompounds_soft_modes}. Notably, $J_1$ is of similar magnitude for the kagome and pyrochlore compounds ($\sim 0.3$~K), and about $3$ and $30$ times smaller for the garnet and the \emph{fcc} compounds, respectively. The smaller magnitude of $J_1$ for the \emph{fcc} lattice is consistent with the minimal antiferromagnetic deviations in the dimensionless susceptibility. The dipolar interaction $D$ in the kagome, garnet, and pyrochlore lattices is around $50$~mK, and much smaller in the \emph{fcc} lattice, $\sim 10$~mK, due to the larger distance between nearest neighbors. The pyrochlore and kagome compounds have the lowest $D/J_1$ ratio, around $0.2$, while the garnet has $D/J_1 \sim 0.4$ and the \emph{fcc} lattice has $D/J_1 \sim 0.9$. Hence the measured $nn$ coupling from Curie Weiss for Ba$_2$GdSbO$_6$ may contain contributions from dipolar interactions. For all subsequent analysis, we consider Ba$_2$GdSbO$_6$ to be a weakly interacting antiferromagnet as discussed in Section~\ref{section:discussion}. 

\begin{table*}[htbp]
\caption{Estimated $nn$ superexchange $J_1 = \frac{3|\Theta|}{zS(S+1)}$ and dipolar interaction $D= \frac{\mu_0 g^2 \mu_B^2}{4\pi R_{nn}^3 k_B}$ for select Gd-based frustrated magnetocaloric materials. The superexchange constant $J_1$ was determined from Curie-Weiss fits of the measured ZFC ($1000$~Oe) magnetic susceptibility $\chi$ from $8-50$~K. Comparisons to the literature-reported values for each compound are also provided. $J_{1,p}$ and $D_p$ refer to the $nn$ superexchange and dipolar term reported for the pyrochlore Gd$_2$Sn$_2$O$_7$~\cite{bondah-jagalu_magnetic_2001,Paddison_Gd2Sn2O7_2017}. The lowest temperature of validity of the classical soft modes theory discussed in the main text is $T^{*} \approx J_1 S$.}
\centering
\label{table:J1_D_Gd_soft_modes}
\begin{ruledtabular}
\begin{tabular}{l c c c c c c c}

\multirow{2}{*}{ } & \multicolumn{2}{c}{Kagome} & \multicolumn{2}{c}{Pyrochlore}  & \multicolumn{2}{c}{Garnet} & \emph{fcc} \\

\multirow{2}{*}{ } & \multicolumn{2}{c}{Gd$_3$Mg$_2$Sb$_3$O$_{14}$} & \multicolumn{2}{c}{Gd$_2$Sn$_2$O$_7$}  & \multicolumn{2}{c}{Gd$_3$Ga$_5$O$_{12}$} & Ba$_2$GdSbO$_{6}$ \\

 & \multicolumn{2}{c}{($z = 4$)} & \multicolumn{2}{c}{($z = 6$)} & \multicolumn{2}{c}{($z = 4$)}  & ($z = 12$)  \\
 

  & Ref. \cite{Wellm_GMSO_2020,Dun_kagome_2017} & $\chi^{-1}$ Fit  & Ref. \cite{bondah-jagalu_magnetic_2001,Paddison_Gd2Sn2O7_2017} & $\chi^{-1}$ Fit & Ref. \cite{Sackville_Hamilton_2014,Paddison_GGG_2015} & $\chi^{-1}$ Fit & $\chi^{-1}$ Fit \\ 
\hline

$\Theta$ (K) & -6(1), -6.70 & -6.8(1) & -9.6(1) & -8.6(1) & -2.6(1) & -2.30(1) & -0.78(1) \\

$J_1$ (K) & 0.3, 0.32  & 0.324(5) & 0.3 & 0.273(3) & 0.107 & 0.110(1) & 0.0124(2) \\

$J_1/J_{1,\textrm{p}}$ & 1 & 1.1 & 1 & 0.9 & 0.4 & 0.4 & 0.04 \\

$D$ (K) & \multicolumn{2}{c}{0.0502} & \multicolumn{2}{c}{0.0496} & \multicolumn{2}{c}{0.0457} & 0.0116 \\

$D/D_{\textrm{p}}$ & \multicolumn{2}{c}{1} & \multicolumn{2}{c}{1} & \multicolumn{2}{c}{0.9} & 0.2 \\

$D/J_1$ & \multicolumn{2}{c}{0.16-0.17} & \multicolumn{2}{c}{0.17} & \multicolumn{2}{c}{0.43} & 0.94\\ 

$T^*$ (K) & \multicolumn{2}{c}{1} & \multicolumn{2}{c}{1} & \multicolumn{2}{c}{0.4} & 0.04 \\

\end{tabular}
\end{ruledtabular}
\end{table*}

\begin{figure}[b]
    \centering
    \includegraphics[width=\columnwidth,clip,trim={0 0 0 0}]{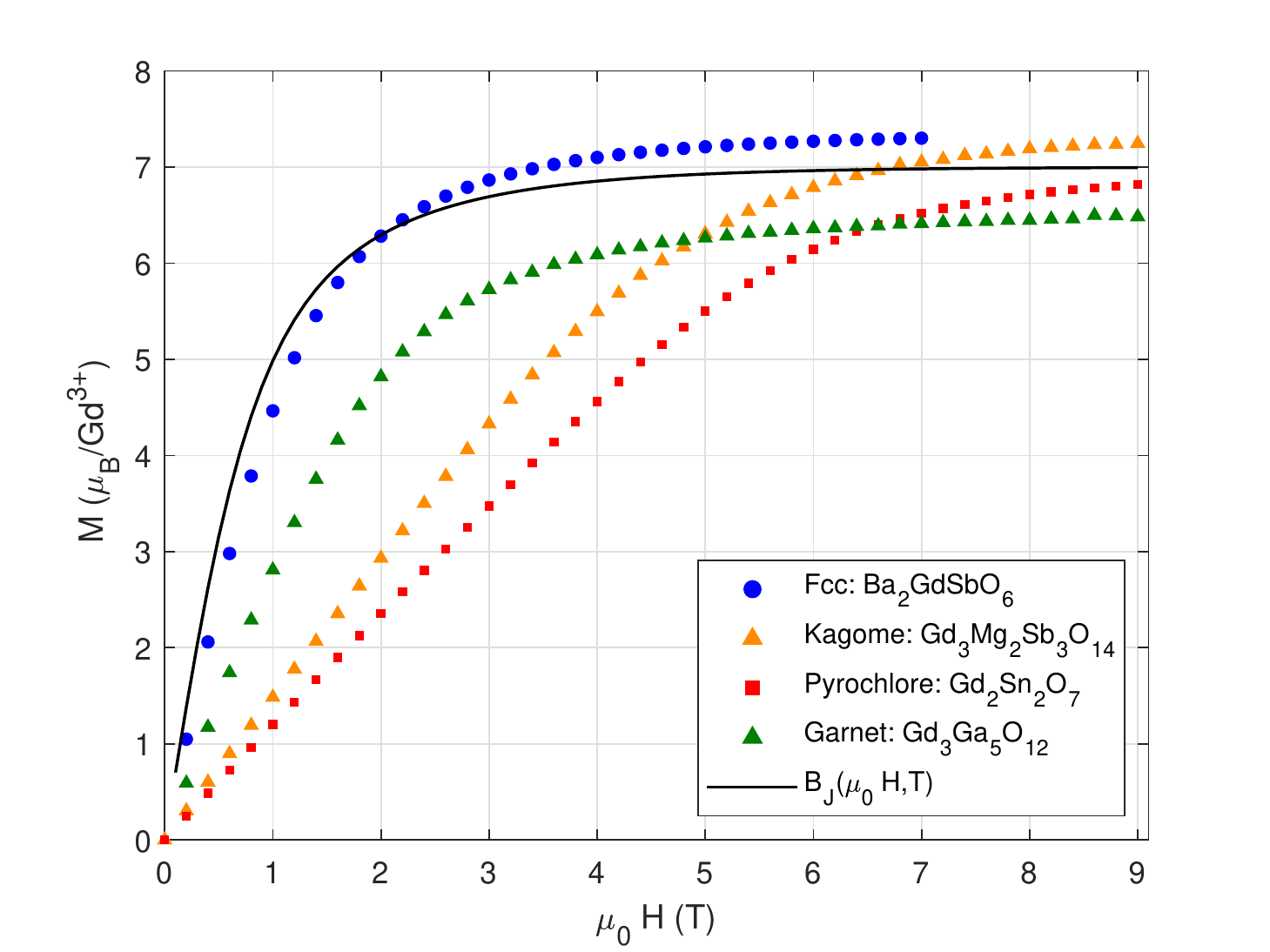}
    \caption{Isothermal magnetization at $2$~K versus applied field $\mu_0 H$ of Gd-based kagome (Gd$_3$Mg$_2$Sb$_3$O$_{14}$), pyrochlore (Gd$_2$Sn$_2$O$_7$), garnet (Gd$_3$Ga$_5$O$_{12}$), and \emph{fcc} (Ba$_2$GdSbO$_6$) compounds. The black solid line gives the theoretically predicted behaviour for uncoupled Heisenberg spins with $S = 7/2$.}
    \label{fig:MvH_Gd_softmodes_2K}
\end{figure}

The isothermal magnetization $M(H)$ was measured at $2$~K, see Figure~\ref{fig:MvH_Gd_softmodes_2K}. Within a field of $9$~T, all compounds reach a maximum value around the $7 \mu_{B}$/${\textrm{Gd}^{3+}}$ predicted for uncoupled Gd$^{3+}$ Heisenberg spins. The \emph{fcc} compound exhibits an isothermal magnetization that agrees well with the uncoupled Heisenberg spins prediction, owing to its small, $\sim 10$~mK superexchange and dipolar interactions, as shown in our prior work~\cite{Koskelo_2022}.

\begin{figure}[htbp]
    \centering
    \includegraphics[width=\columnwidth,clip,trim={0 0 0 30}]{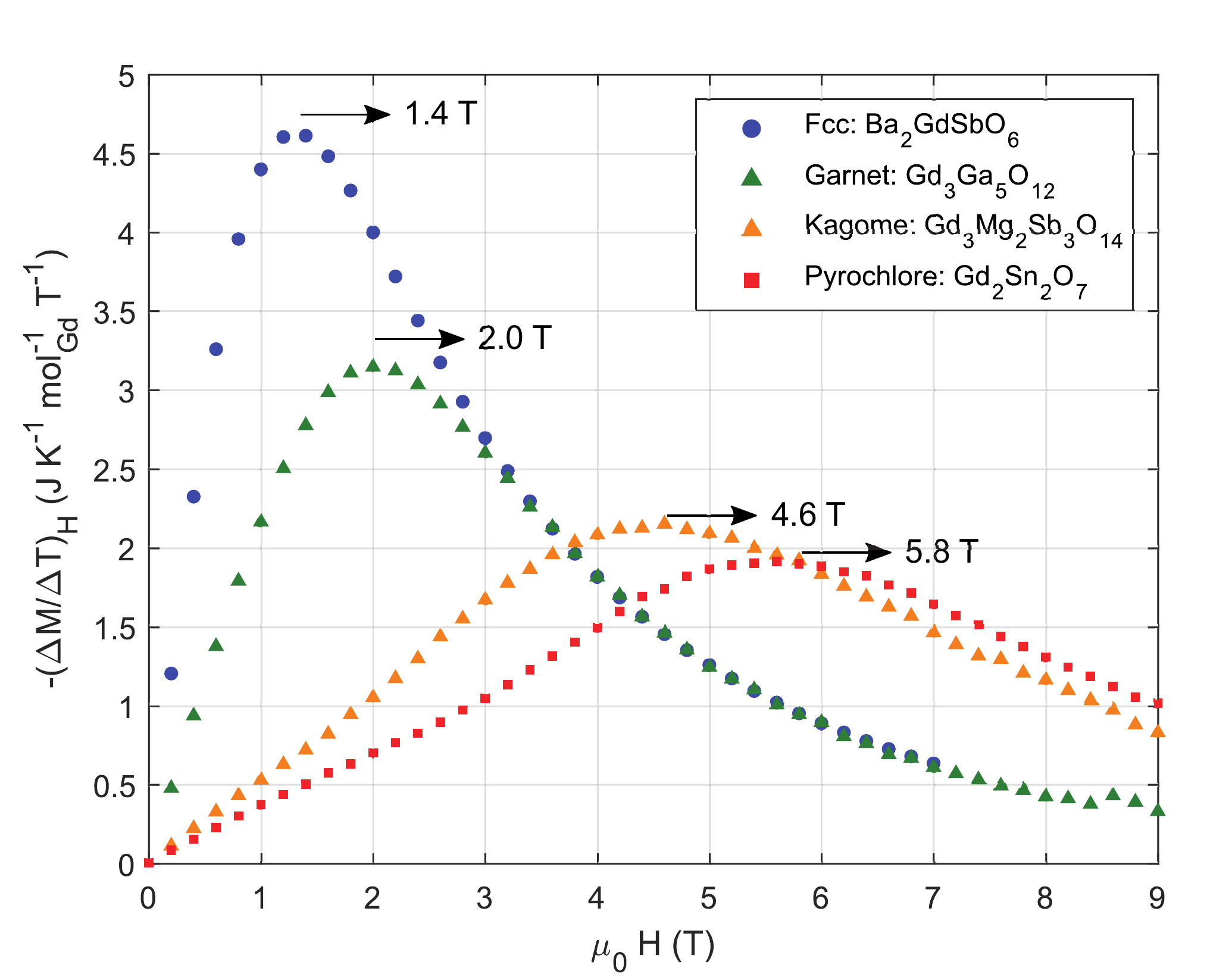}
    \caption{Approximate magnetocaloric cooling rate $-(\Delta M/\Delta T)_H \approx -(\partial M/\partial T)_H$ versus field for Ba$_2$GdSbO$_6$, Gd$_3$Ga$_5$O$_{12}$, Gd$_3$Mg$_2$Sb$_3$O$_{14}$, and Gd$_2$Sn$_2$O$_7$ at $2$~K. These were obtained from experimental measurements of isothermal magnetization with $\Delta T = 2$~K. The resulting value of the saturation field $H_{\rm sat}$ for each material is indicated by the arrows.}
    \label{fig:dMdT_at2K_Gdcompounds_Hsat}
\end{figure}

The saturation field of each compound was estimated from the field at which the temperature gradient of the magnetization, $-(\partial M/\partial T)_H = -(\partial S_m/\partial H)_T$ shown in Figure~\ref{fig:dMdT_at2K_Gdcompounds_Hsat}, is maximized~\cite{Zhitomirsky_2003,Gd2Ti2O7_MCE_Tad_Zhitomirksy}. This method of estimating the saturation field should be reliable in the temperature range where spin correlations are relevant, i.e. for $T \lesssim J_1 S (S+1)$. Theoretical predictions for $H_{\rm sat}$ (at zero temperature) based on the large-$S$ Heisenberg Hamiltonian, Equation~\eqref{eq:spinH}, are listed in Table~\ref{table:Hsat_Gd_compounds_soft_modes}. The observed values for the corner-sharing geometries agree well with the predictions, indicating that the compounds are well described by the Heisenberg model and suggests that for $T\geq2$~K, $nn$ superexchange $J_1$ plays a key role in the magnetocaloric cooling rate, $(\partial S_m/\partial H)_T$. As described in Section~\ref{section:discussion}, the \emph{fcc} compound Ba$_2$GdSbO$_6$ can best be described as a weakly frustrated Heisenberg antiferromagnet. At $2$~K, its magnetocaloric cooling rate can be modeled well using a mean-field superexchange model, in which spatial spin correlations are altogether neglected. 
Due to its small superexchange, $J_1S(S+1) \sim$ 0.2 K, lower temperature $M(H)$ measurements are needed to measure the saturation field with our experimental protocol. This is beyond the scope of the present work and, for completeness we report here the estimate of the saturation field at $T = 2$ K, the lowest temperature measured. 

%
%

\subsection{Magnetocaloric Effect}

From one of Maxwell's relations, the isothermal field gradient of the magnetic entropy is related to the temperature gradient of the magnetization at constant field via $\left(\frac{\partial S_m}{\partial H} \right)_T = \left( \frac{\partial M}{\partial T} \right)_H $. Thus, the magnetic entropy change, $\Delta S_m$, can be measured using the isothermal magnetization via:
\begin{equation}
    \Delta S_m(T_0, H_{\rm max}) = \int_0^{H_{\rm max}} \left.\left(\frac{\partial M(T,H)}{\partial T} \right)_H \right \vert_{T = T_0} dH \, .
    \label{eq:deltaSm_fromdMdT}
\end{equation}


\begin{figure*}[htbp]
    \centering
    \includegraphics[width=1.95\columnwidth,clip,trim = {70 0 20 0}]{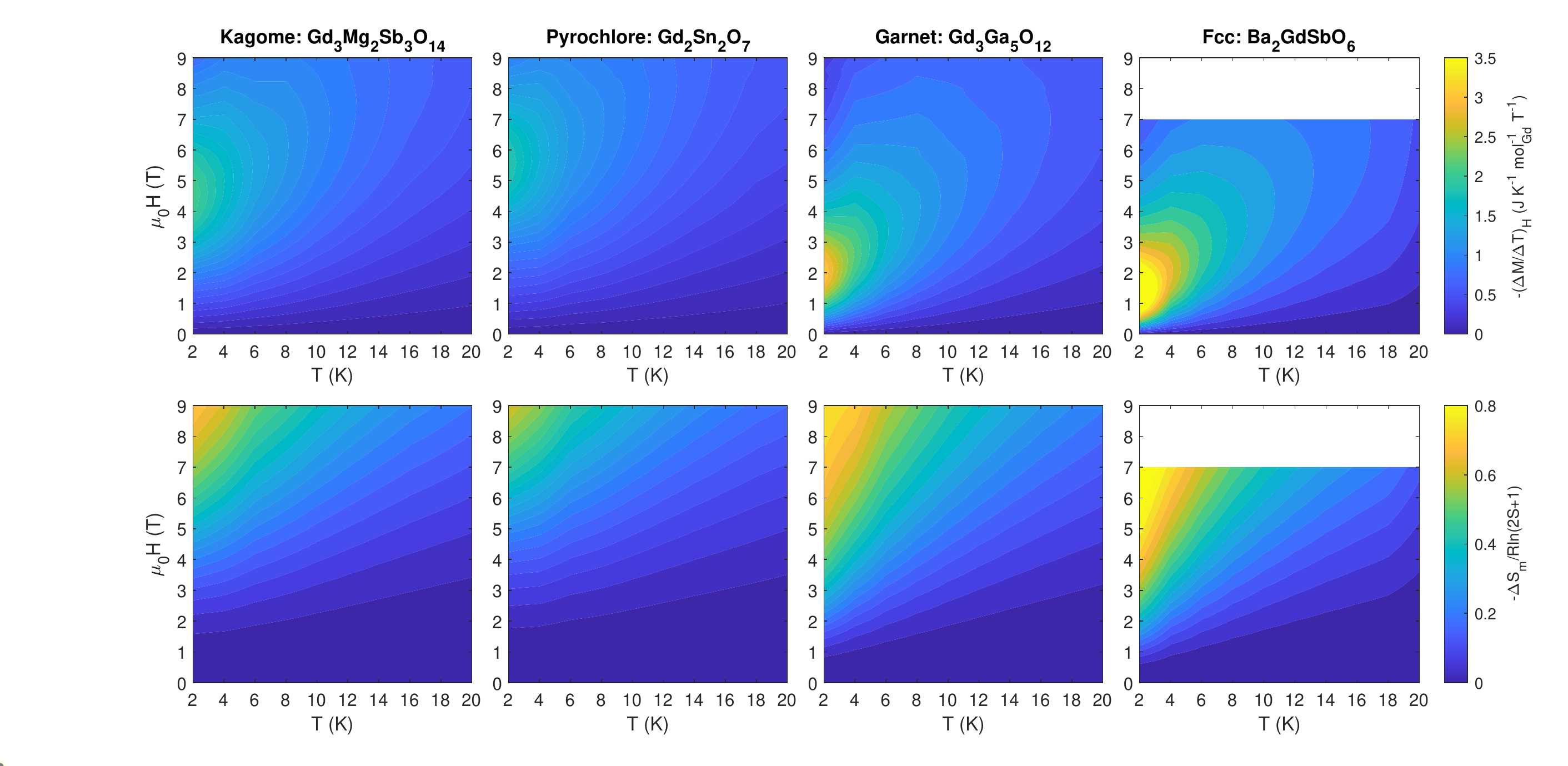}
    \caption{Measured temperature gradient of the magnetization $-(\Delta M/\Delta T)_H \approx -(\partial M/\partial T)_H$ (with $\Delta T = 2$~K) and resulting magnetic entropy change $\Delta S_m$, normalized by the maximum free-spin value, $R\ln(2S+1)$, from $2-20$~K under applied fields of $0-9$~T for the Gd-based frustrated magnetocaloric materials discussed in this work.}
    \label{fig:deltaSm_soft_modes_noNorm}
\end{figure*}

The magnetic entropy change $\Delta S_m$ of each Gd-based compound was measured from $2$ to $20$~K in fields of up to $9$~T from the isothermal magnetization, Figures~\ref{fig:deltaSm_soft_modes_noNorm} and~\ref{fig:deltaSm_slices}a). We note that the magnetization data were measured in temperature steps of $2$~K, and thus we are only able to measure the finite difference ratio $(\Delta M/\Delta T)_H$ as an estimate for the magnetocaloric cooling rate $(\partial M/\partial T)_H$, prior to computing the magnetic entropy change via Equation~\ref{eq:deltaSm_fromdMdT} (see Appendix \ref{app:methods}). 

Per mol Gd, the maximum entropy change attainable for a paramagnetic salt is $R\ln(2S+1)$ ($17.28$~J~K$^{-1}$~mol$^{-1}_{\rm Gd}$). We find that the \emph{fcc} Ba$_2$GdSbO$_6$ exhibits the greatest magnetic entropy change $-\Delta S_m$, reaching $0.9 R\ln(2S+1)$ in a field of just $7$~T at $2$~K. On the other hand at $2$~K and a larger field of $9$~T, the corner-sharing geometries each exhibit entropy changes of $0.8$, $0.7$, and $0.6 R\ln(2S+1)$ for the garnet, kagome, and pyrochlore lattices, respectively. At a low field of $2$~T, the \emph{fcc} and garnet compounds are still the best performing with $-\Delta S_m$ of $0.4 R\ln(2S+1)$ and $0.2 R\ln(2S+1)$, respectively, compared to $\sim0.05 R\ln(2S+1)$ for the kagome and pyrochlore, Figure~\ref{fig:deltaSm_slices}a). Despite having the largest number of soft modes, the pyrochlore material achieves the smallest magnetic entropy change per Gd$^{3+}$ ion. 


\section{Discussion}
\label{section:discussion}

\begin{figure}
    \centering
    \includegraphics[width=\columnwidth,clip,trim={10 35 10 30}]{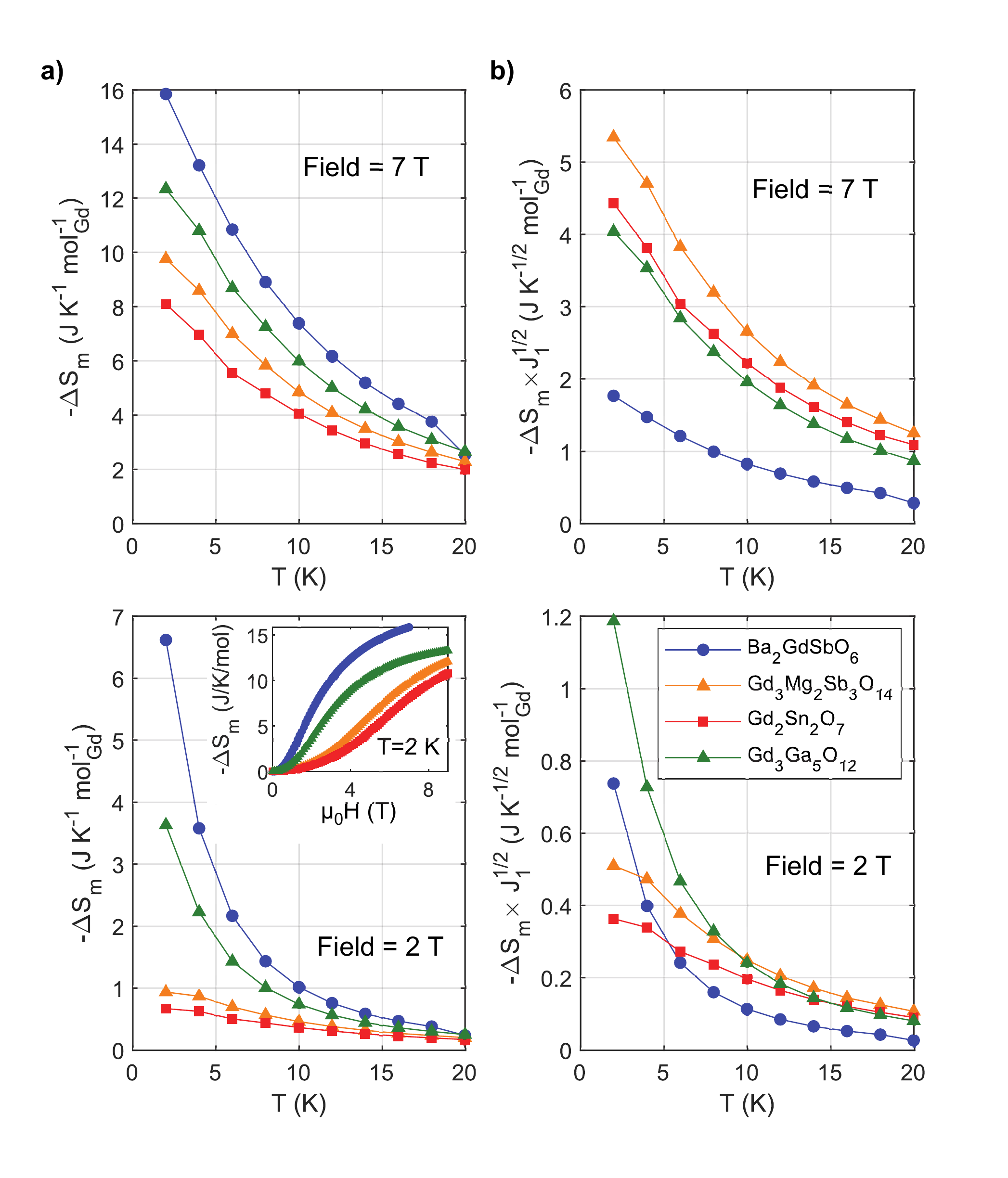}
    \caption{a) Magnetic entropy change $\Delta S_m$ per mol$_\textrm{Gd}$ versus temperature at fields of $7$~T and $2$~T (inset: $\Delta S_m$ versus field at $2$~K). b) Superexchange-normalized magnetic entropy change $\Delta S_m \times \sqrt{J_1}$ versus temperature at fields of $7$~T and $2$~T.}
    \label{fig:deltaSm_slices}
\end{figure}

The enhanced performance of the \emph{fcc} Ba$_2$GdSbO$_6$ and garnet Gd$_3$Ga$_5$O$_{12}$ compared to the pyrochlore Gd$_2$Sn$_2$O$_7$ is at first surprising, given the reduced number of soft modes (Table~\ref{table:Hsat_Gd_compounds_soft_modes}). 
However, from Equations~\eqref{eq:sing_Zhito} and~\eqref{eq:N2_modes_Zhito} we see that the measured magnetocaloric cooling rate results from two different contributions, one due to the number of soft modes $N_4$ (frustrated response) and one due to ordinary dispersive modes $N_2$ (paramagnetic response). 
In order to investigate this in greater detail, we compare the experimental data with mean-field modelling~\cite{Koskelo_2022} and classical Heisenberg Monte Carlo simulations of the finite difference ratio $-(\Delta M/\Delta T)_H$ for each of the four systems (see Figure~\ref{fig:4pane_J1fits_dMdT}). 
For reference, we show in Figure~\ref{fig:dSdH_MCvMFT_appendix} in the Appendix a comparison between the finite difference ratio $-(\Delta M/\Delta T)_H$ and the exact magnetocaloric effect $-(\partial S_m/\partial H )_T = -(\partial M/\partial T )_H$, for both mean-field and Monte Carlo results. 


Our results show that the \emph{fcc} compound is described quantitatively well by the mean-field $nn$ superexchange model~\cite{Koskelo_2022} across the full range of temperatures of interest in this study~\footnote{The progressive discrepancy observed between experiment and Monte Carlo simulations at large applied fields is likely due to spin quantization effects.}. This finding suggests that the paramagnetic response dominates for this compound in the temperature and field regime of interest. 
We recall indeed that the \emph{fcc} lattice is predicted to only have a subextensive number of soft modes, and therefore the contribution $\left(\frac{\partial S_m}{\partial H} \right)_{T,N4}$ at the saturation field becomes irrelevant in the thermodynamic limit. The paramagnetic term $\left(\frac{\partial S_m}{\partial H} \right)_{T,N2}$ is expected to be the dominant contribution to the measured magnetic entropy change $\Delta S_m$. 



In contrast, for the other three (corner-sharing) compounds we clearly see an increasing discrepancy between mean-field and experiment at lower temperatures ($2$ and $4$~K). This is consistent with the $10$-$30$-fold larger values of $J_1$ in these systems, resulting in correspondingly stronger correlations. 
The latter are generally expected to have two effects: (i) reduce the $N_2$ contribution to the magnetocaloric effect (the spins have reduced ability to fluctuate independently of one another); and (ii) give rise to an extensive number of frustrated collective soft mode $N_4$, which contribute with a prefactor scaling as $1/\sqrt{T}$ to $(\partial S_m/\partial H )_T$. 
The comparatively better agreement of the experimental curves with classical Monte Carlo simulations, with respect to mean-field, demonstrates how the latter are able to capture both the paramagnetic as well as the soft modes contribution, with corrections at large fields due to quantization effects~\footnote{We notice a discrepancy between Monte Carlo simulations and experiments at small field values and low temperatures in the kagome case, whose origin remains presently unclear.}. 

%

\begin{figure*}
    \centering
    \includegraphics[width = 1.8\columnwidth,clip,trim={10 35 10 20}]{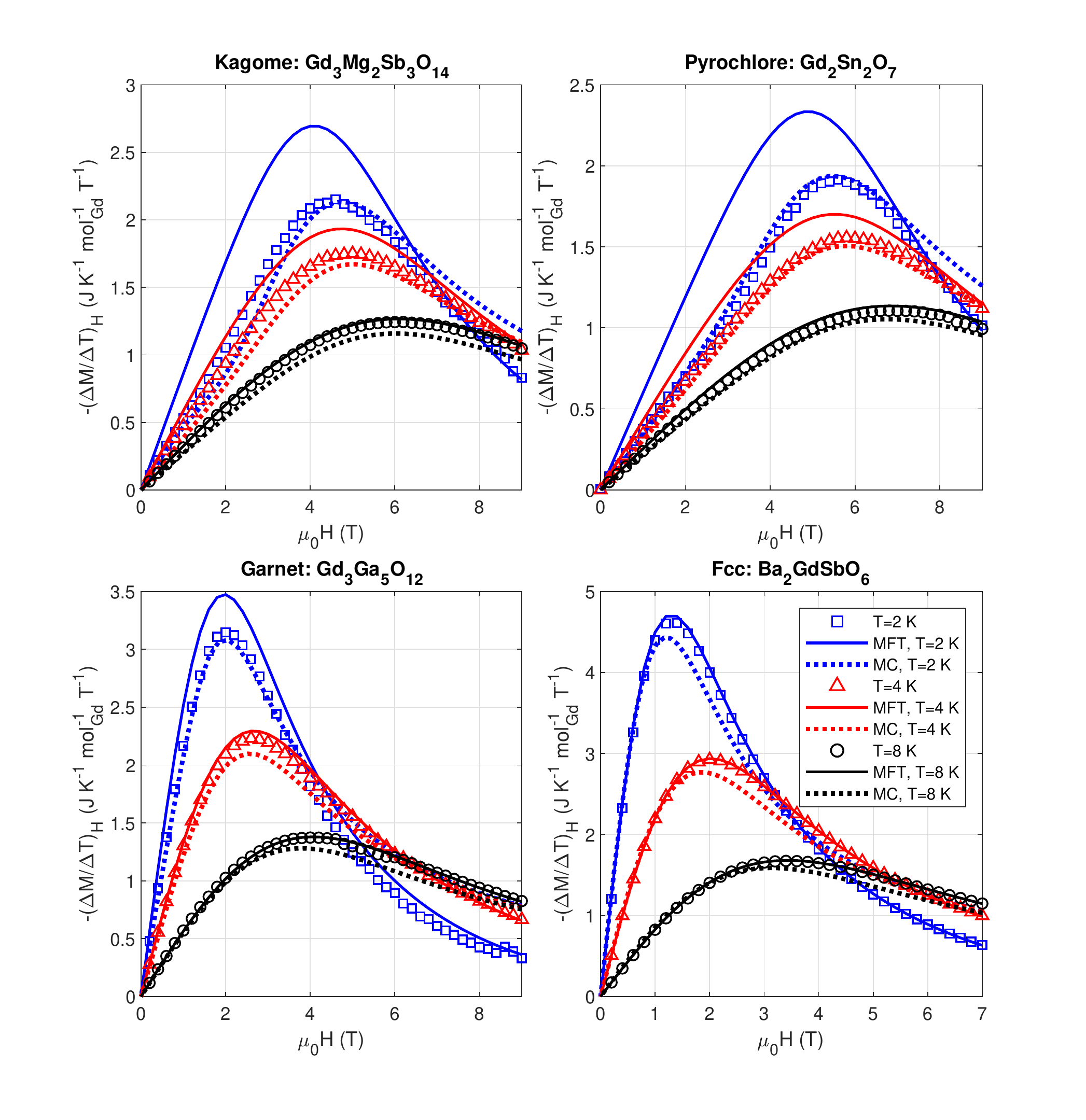}
    \caption{Approximate magnetocaloric cooling rate $-(\Delta M/\Delta T)_H \approx -(\partial M/\partial T)_H = -(\partial S_m/\partial H)_T$ extracted from the measured isothermal magnetization with $\Delta T = 2$~K (data points), compared to predictions based on the magnetization $M$ for Heisenberg $S = 7/2$ spins in the mean-field superexchange model (MFT)
    (solid lines) and classical Monte Carlo simulations (MC) (dotted lines). These results show that the \emph{fcc} compound is well described by mean-field quantum spins, which neglect spatial correlations. Conversely, the better agreement of classical Monte Carlo for the corner-sharing geometries highlights the importance of magnetic correlations in those compounds.}
    \label{fig:4pane_J1fits_dMdT}
\end{figure*}

\begin{figure*}[htbp]
    \centering
    \includegraphics[width=1.95\columnwidth,clip,trim={60 0 0 0}]{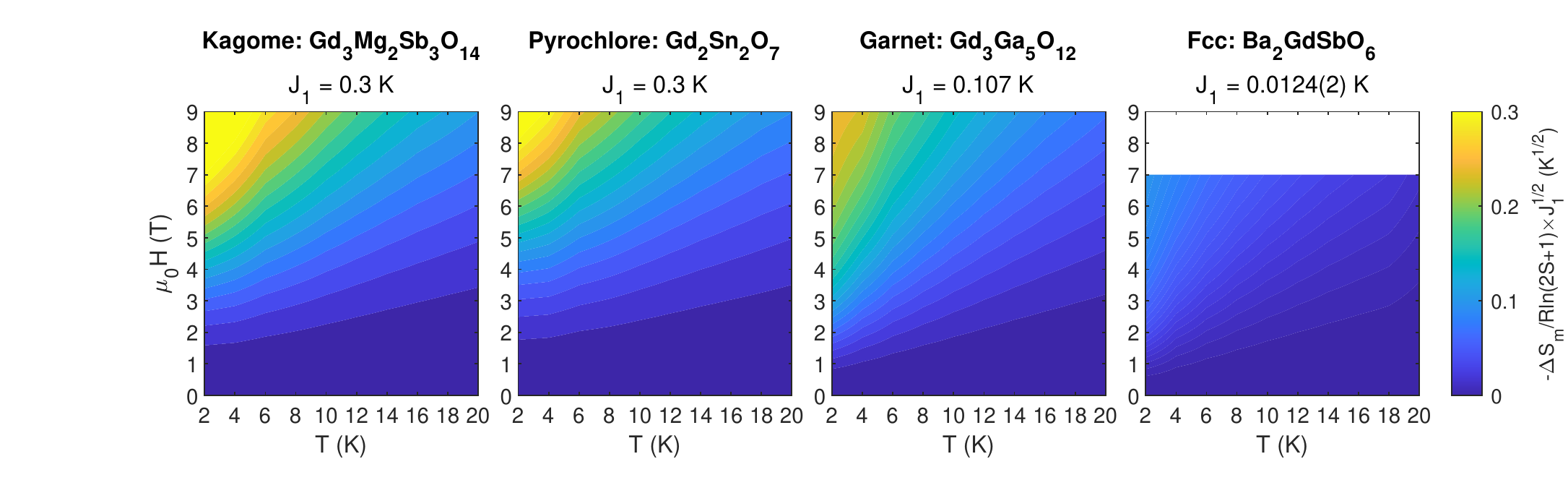}
    \caption{Measured magnetic entropy change normalized by the $nn$ superexchange, $\Delta S_m \times \sqrt{J_1}$. Accounting for the differences in the superexchange term, the pyrochlore lattice becomes the highest performing among the 3D frustrated lattices.}
    \label{fig:deltaSm_soft_modes_J1Norm}
\end{figure*}

Having established the importance of soft modes in the magnetocaloric performance of the corner-sharing geometries, it remains to be seen why the garnet outperforms both the pyrochlore and the kagome lattices.
This can be explained by the fact that the measured entropy changes in Figures~\ref{fig:deltaSm_soft_modes_noNorm} and~\ref{fig:deltaSm_slices}a) do not account for differences in the value of the superexchange coupling $J_1$. To account for the role of superexchange in the predicted enhancement at the saturation field, Equation~\eqref{eq:sing_Zhito}, the magnetic entropy maps were scaled by $\sqrt{J_1}$, Figures~\ref{fig:deltaSm_slices}b) and~\ref{fig:deltaSm_soft_modes_J1Norm}. While the soft modes contribution to the magnetic entropy change is expected to vanish for the \emph{fcc} lattice due to its subextensive $N_4$, we include the Ba$_2$GdSbO$_6$ compound for ease of comparison. After normalization, the pyrochlore lattice becomes the top-performer among the 3D geometries at $7$~T, as predicted by Ref.~\onlinecite{Zhitomirsky_2003}, while the kagome sample is the top-performer overall.

The scaled magnetic entropy change, $-\sqrt{J_1} \Delta S_m$, agrees qualitatively with the soft modes model. From Equation~\eqref{eq:sing_Zhito}, the maximum in $(\partial S_m/\partial H)_T$ at the saturation field $H_{\rm sat}$ should scale proportionally to the number of soft modes $N_4$, once normalized by the square root of the superexchange $\sqrt{J_1}$. This qualitative agreement is striking as it reproduces the predicted ranking for the three-dimensional frustrated lattices of the pyrochlore ($N_4 \sim N$), followed by the garnet ($N_4 \sim 2N/3$), and then the \emph{fcc} ($N_4 \sim N^{1/3}$) for applied fields greater than or equal to saturation (i.e., $H \geq$ $5.8$~T). 
At the low field of $2$~T, the kagome and pyrochlore compounds are not fully saturated, explaining why, as might be expected, the garnet and \emph{fcc} lattices exhibit the greatest $-\sqrt{J_1} \Delta S_m$, Figure~\ref{fig:deltaSm_slices}b). A more quantitative assessment of the agreement with the model at lower fields is not possible, likely due to other minor contributions (e.g., dipole-dipole interactions and disorder), which were not accounted for in our study. 

The implication of this experimental validation is that while the order of soft modes of frustrating lattices can be accurately captured in experiment, soft modes are not the leading contribution in the magnetocaloric performance of a frustrated magnet in the liquid He regime. Rather, as demonstrated in Figures~\ref{fig:deltaSm_soft_modes_noNorm} and~\ref{fig:deltaSm_slices}, materials with the smallest superexchange (i.e., the \emph{fcc} and garnet) exhibit the greatest performance, as the paramagnetic contribution to the entropy change is not constrained by short-range correlations so that spins can fluctuate independently of one another. It is conceivable that at lower temperatures relative to the superexchange $J_1$ ($\ll 2$~K), soft modes may become the determining factor in the magnetocaloric effect. However, such low temperatures would soon require a revision of the model to include quantum statistics to study spin excitations. 

These results suggest that future research for cryogenic magnetic refrigeration should focus on frustrated magnets with the smallest superexchange. The strategy of solely reducing the exchange by increasing the distance between magnetic ions, as in the dilute magnetic salts FAA and CMN, is not effective. When the four materials studied here are normalized by volume and mass, see Appendix \ref{app:entropypervol}, then the \emph{fcc} lattice is the poorest performer despite having the weakest superexchange coupling. This highlights the importance of reducing $J_1$ while maintaining a dense magnetic lattice \cite{Koskelo_2022}. Furthermore, in these dense lattices, soft modes could provide an additional cooling mechanism at lower temperatures. 

Another interesting result from this study is that, in the limit where superexchange determines the magnetocaloric cooling rate, differences between the pyrochlore and kagome lattices are not predicted by the model. In this case, they may arise due to the presence of site disorder (10.5(2)\%) in Gd$_3$Mg$_2$Sb$_3$O$_{14}$ (Appendix~\ref{appendix:Gd_structures_soft_modes}) relieving frustration, or spin-anisotropies not included in this analysis, which only applies to isotropic Heisenberg spins. 

This study focuses on four frustrated oxides. However, there are many other magnetic lattices where a high magnetocaloric effect is reported including the 2D triangular (GdBO$_3$~\cite{GdBO3_2018}) and Shastry-Sutherland lattices (Gd$_2$Be$_2$GeO$_7$~\cite{Gd2Be2GeO7_2021}), and quasi-1D spin chains (Gd(HCOO)$_3$, GdOHCO$_3$, Ca$_4$GdO(BO$_3$)$_3$~\cite{Gd-formate_2013,GdOHCO3_2014,1D_borates_Nicola}). In many of these systems the presence of polyanions (BO$_3^{3-}$, HCOO$^{-}$, etc.) or molecular ligands may complicate the soft mode analysis presented in Reference~\onlinecite{Zhitomirsky_2003} due to changes in the hydrogen bonding or other molecular interactions~\cite{saines_probing_2018,hellsvik_spin_2020}. Such materials are beyond the scope of this initial study. Future work could incorporate the soft mode analysis of other geometrically frustrated magnetic lattices, e.g. edge-sharing motifs such as the edge-sharing triangles in a honeycomb arrangement in SrLn$_2$O$_4$~\cite{Karunadasa_2005}. 

The mean-field nature of the analysis conducted here is only applicable to Heisenberg spins. Anisotropic magnetic systems such as those based on Dy$^{3+}$ have been found as useful magnetocaloric candidates at low fields (e.g., $\mu_0 H \leq 2$~T), compared to the Gd$^{3+}$ counterparts. For example, the magnetic entropy change of Ising garnet Dy$_3$Ga$_5$O$_{12}$ reaches four times that in the Heisenberg Ga$_3$Ga$_5$O$_{12}$ for a $1$~T field at $2$~K~\cite{DGGvGGG_MCE}. Adiabatic temperature measurements have shown that the Ising spin ice Dy$_2$Ti$_2$O$_7$ can cool from $\sim 1$~K to $0.3$~K in a field of just $0.8$~T, while Gd$_2$Ti$_2$O$_7$ requires a field of $\sim 9$~T to cool to $0.5$~K from the same starting temperature~\cite{Tad_Dy2Ti2O7,Tad_Gd2Ti2O7}. Extending this analysis to anisotropic systems is an interesting avenue for future research, likely requiring non-trivial microscopic modeling of the single-ion anisotropy.
%
%

\section{Conclusion}

This work examined the role of lattice geometry in maximizing the magnetocaloric effect via soft mode spin excitations. Four representative Heisenberg (3D, Gd-based) spin systems were investigated including the three corner-sharing garnet (Gd$_3$Ga$_5$O$_{12}$), pyrochlore (Gd$_2$Sn$_2$O$_7$), kagome lattices (Gd$_3$Mg$_2$Sb$_3$O$_{14}$), and the edge-sharing \emph{fcc} lattice (Ba$_2$GdSbO$_6$). For the liquid-He temperature range investigated ($2-20$~K), magnetic entropy change measurements indicate that the smaller superexchange of the \emph{fcc} and garnet lattices allows for better magnetocaloric performance than the strongly-coupled pyrochlore analog, despite its larger number of soft modes. Our results show that a paramagnetic response dominates above geometric frustration in enhancing the magnetocaloric effect. However, the contribution to the magnetocaloric effect in systems with an extensive number of frustrated magnetic soft modes is expected to take over at lower temperatures, thus providing an additional channel for maximizing the magnetocaloric effect in highly frustrated magnetic lattices. Future magnetocaloric materials design should focus on both the superexchange and magnetic lattice to optimize the trade off between spin polarizability and frustrated soft mode enhancement.
%
%

\begin{acknowledgments}
For the purpose of open access, the author has applied a Creative Commons Attribution (CC BY) licence to any Author Accepted Manuscript version arising from this submission. This work was supported supported by the Engineering and Physical Sciences Research Council (EPSRC) grants (EP/P034616/1, EP/V062654/1 and EP/T028580/1) and the Winton Programme for the Physics of Sustainability. E.C.K. gratefully acknowledges the support of a Churchill Scholarship from the Winston Churchill Foundation of the United States. M.E.Z. acknowledges financial support from Agence Nationale de la Recherche,
France, Grant No. ANR-18-CE05-0023. Magnetic measurements were made on the EPSRC Advanced Characterization Suite EP/M0005/24/1.
\end{acknowledgments}
%
%
\appendix

\section{\label{app:Hsat}Saturation field of the \emph{fcc} model}

One can use frustrated block decomposition  to obtain the saturation field $H_\textrm{sat}$ for the nearest neighbor
\emph{fcc} antiferrromagnet, above which spins become fully aligned with the applied field. We start with the spin Hamiltonian
\begin{equation}
{\cal H} = J_1 \sum_{\langle ij \rangle} \vec{S}_i\cdot\vec{S}_j -  H \sum_i S_i^z \,, 
\label{eq:afmH}
\end{equation}
where the $g \mu_B$ prefactor in the Zeeman term has been absorbed into $H$.
An \emph{fcc} lattice can be represented as an edge-sharing arrangement of tetrahedra, which are labeled by an index $\alpha$. Introducing the tetrahedron magnetization $\vec{L}_\alpha = \sum_{i \in \alpha} \vec{S}_i$, and using the fact that every site belongs to $8$ tetrahedra, it can be shown that:
\begin{equation}
\sum_i S_i^z = \frac{1}{8} \sum_\alpha L_\alpha^z \,.
\end{equation}
Similarly, since every $nn$ bond of the \emph{fcc} lattice belongs to $2$ tetrahedra, we find that:
\begin{equation}
\sum_{\langle ij \rangle} \vec{S}_i\cdot\vec{S}_j = \frac{1}{4} \sum_\alpha \bigl( \vec{L}_\alpha \bigr)^2 + {\rm const.} 
\end{equation}

The Hamiltonian~\eqref{eq:afmH} can be now rewritten, up to an irrelevant constant, as: 
\begin{eqnarray}
{\cal H} &=& \frac{1}{4} \sum_{\alpha} \left[ J_1 \vec{L}_\alpha^2 - \frac{H}{2} L_\alpha^z \right] 
\nonumber \\
&=& \frac{J_1}{4} \sum_{\alpha} \left[ \vec{L}_\alpha - \frac{H \hat{z}}{4 J_1} \right]^2 + {\rm const.}
\, . 
\label{eq:afmH_L}
\end{eqnarray}
For a single tetrahedron, the energy is minimized when $\vec{L}_\alpha = (H/4J_1) \hat{z}$. If this condition can be met for \emph{all tetrahedra simultaneously}, then it is the lowest energy state of the system and its zero-temperature state. 

Because $\vert \vec{S}_i \vert = S$,  $L_\alpha^z$ is upper bounded by the maximum possible value $(L_\alpha^z)_{\rm max} = 4S$, if $H > 16 J_1S$ then the minimum energy condition cannot be achieved and the lowest energy
state corresponds the uniformly polarized  spin configuration. This value is referred to as the saturation field, 
\begin{equation}
g\mu_B H_{\rm sat} = 16 J_1S \, .
\label{eq:Hsat}
\end{equation}
%
%

\section{\label{app:meanfield}
Mean-field Approximation}

A mean-field approximation (random phase approximation)~\cite{Wellm_GMSO_2020,Koskelo_2022} was used to estimate the paramagnetic contribution to the magnetocaloric cooling rate $(\partial S_m/\partial H)_T$ for each compound, truncating the interactions at nearest neighbour distance. The antiferromagnetic coupling between $S=7/2$ spins amounts to an exchange field, $\vec{H}_{\rm exc}$, which adds to the external field, $\vec{H}_{\rm ext}$, to produce the net field, $\vec{H}_{\rm tot}$, experienced by a single spin. 

Note that dipolar interactions can be neglected for the corner sharing lattice compounds considered in this study since they are weak in comparison to the nearest neighbor exchange (see Table~\ref{table:J1_D_Gd_soft_modes}). This is not the case for the \emph{fcc} material. However, the dipolar interactions contribution for a cubic Bravais lattice vanishes identically in the case of a (uniform) mean-field approximation, due to rotational lattice symmetry. 

Since $L = 0$ for Gd$^{3+}$, the exchange constant $J_1$ can be assumed to be isotropic, so that the exchange field is given by:
\begin{equation}
    \vec{H}_{\rm exc} = a_{\rm ex} z M \hat{H}_{\rm ext} = \frac{-J_1}{g^2 \mu_B} z M \hat{H}_{\rm ext} \, ,
    \label{eq:Wellm_Hexc}
\end{equation}
where $M$ is the bulk magnetization in units of the Bohr magneton and $a_{\rm ex}$ is the ``field parameter'' in units of magnetic field~\cite{Wellm_GMSO_2020}. The bulk magnetization of the system at a given temperature $T$ and external field $H_{\rm ext}$ is given by the solutions of the transcendental self-consistency equation:
\begin{equation}
    M - g S \, B_S\left( \left\vert \vec{H}_{\rm tot}(M) \right\vert,T \right) = 0 \, ,
    \label{eq:Wellm_trans_eq_M}
\end{equation}
where $B_S$ is the Brillouin function given by: 
\begin{equation}
B_S(y) = \frac{2S + 1}{2S}\coth\left(\frac{2S + 1}{2S} y\right) - \frac{1}{2S}\coth\left(\frac{y}{2S}\right) \, ,
\label{eq:B_Heis_intro}
\end{equation} 
with $y = \mu_0 H g \mu_B S / (k_B T)$, and $H = \vert \vec{H}_{\rm tot} \vert$~\cite{Wellm_GMSO_2020}. 

\begin{table}[b]
    \centering
     \caption{Mean-field saturation magnetization parameter $M_{sat}$ determined by fitting the observed saturation magnetization, as well as the $nn$ superexchange $J_1$, for each compound, and $R^2$ of the fit.}
    \begin{ruledtabular}
    \begin{tabular}{l|c|c|c}
     Compound & $M_{sat}$ ($gS$) & Fit $J_1$ (K) & $R^2$ \\
     \hline
     Gd$_3$Mg$_2$Sb$_3$O$_{14}$  & 1.0268 & 0.312(3) & 0.9982 \\
     Gd$_2$Sn$_2$O$_7$ &  0.9744 & 0.273(2) & 0.9976  \\
     Gd$_3$Ga$_5$O$_{12}$ &   0.9257 & 0.114(1) & 0.9997 \\
     Ba$_2$GdSbO$_6$ &   1.04      & 0.0113(3)     & 1.0000 \\
    \end{tabular}
    \end{ruledtabular}
    \label{table:mean-field-fits}
\end{table}

As a further consistency check we leave $J_1$ as a free parameter in our model. Global least-squares fits to the measured isothermal magnetization $M(H)$ ($2-20$~K) using Equation~\eqref{eq:Wellm_trans_eq_M} were used to determine the $nn$ superexchange $J_1$ in Ba$_2$GdSbO$_6$, Gd$_2$Sn$_2$O$_7$, Gd$_3$Ga$_5$O$_{12}$ and Gd$_3$Mg$_2$Sb$_3$O$_{14}$. The fit $J_1$ values, Table~\ref{table:mean-field-fits}, are in agreement with the literature reported values and Curie-Weiss fits, Table~\ref{table:J1_D_Gd_soft_modes}. The free-spin magnetization $M_{S=7/2} = g S B_S (|\vec{H}_{\rm ext}|,T)$ was used as an initial guess to solve iteratively the mean-field self-consistency condition, Equation~\eqref{eq:Wellm_trans_eq_M}. The observed saturation value of the magnetization at the maximum field can vary due to experimental uncertainty, so all compounds were fit with a scaled fraction of $M_{\rm sat} = g S$, to match the observed value, see Table~\ref{table:mean-field-fits}. 

\begin{figure*}[htbp]
    \centering\includegraphics[width=1.5\columnwidth,clip,trim={30 135 40 120}]{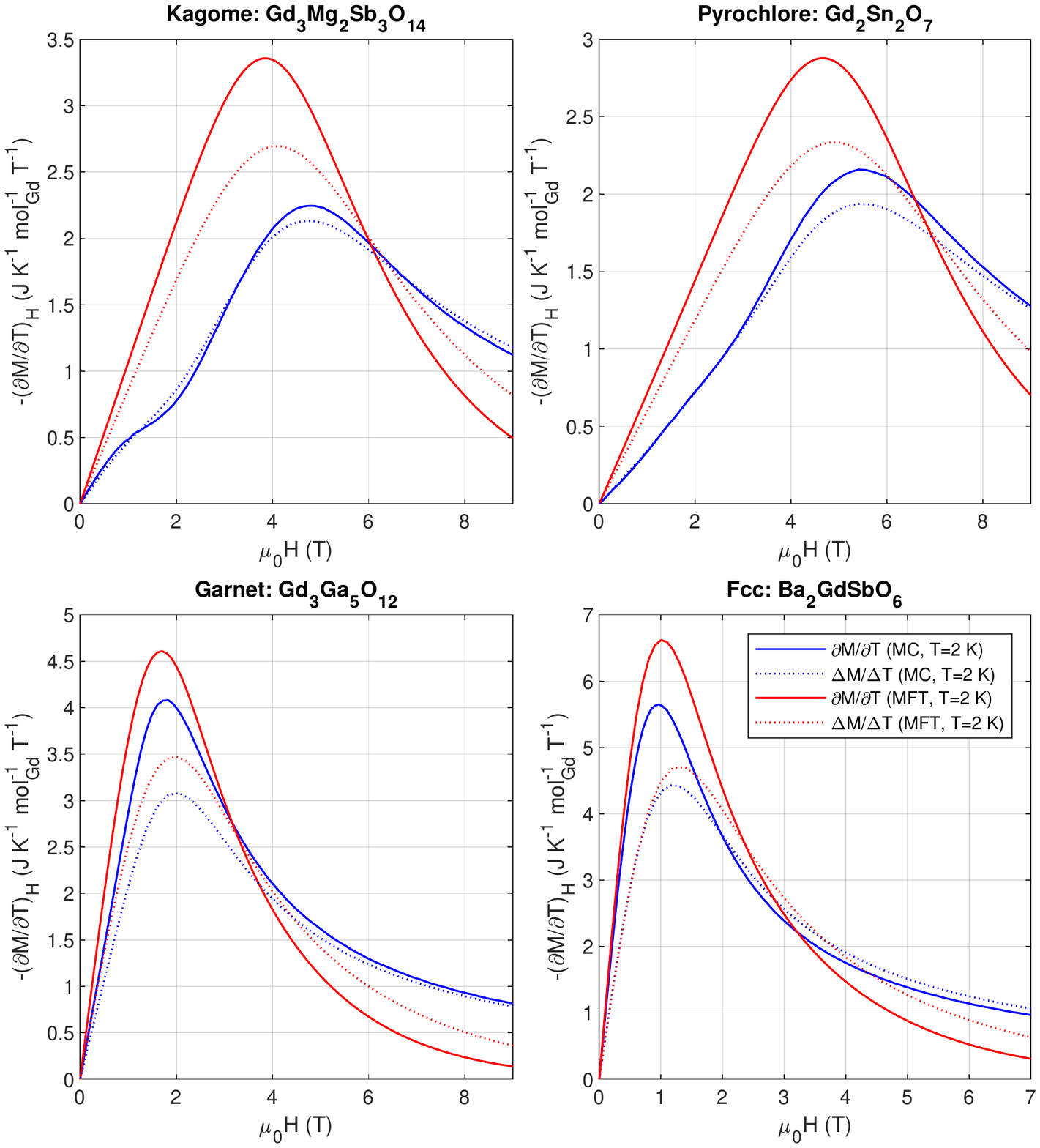}
    \caption{Comparison of the exact,  $-(\partial S_m/\partial H)_T = -(\partial M/\partial T)_H$, vs approximate, $-(\Delta M/\Delta T)_H$ with $\Delta T=2$~K, magnetocaloric cooling rates for the four frustrating geometries, using classical Monte Carlo (MC) (blue) and mean-field theory (MFT) (red) at T = 2 K. Exact magnetocaloric cooling rates are shown as solid lines while approximate rates are shown as dotted lines.}
    \label{fig:dSdH_MCvMFT_appendix}
\end{figure*}

The approximate magnetocaloric cooling rate $(\Delta M/\Delta T)_H$ was then calculated from the model predictions of $M(H)$ using the fit $nn$ superexchange constants $J_1$, Equation~\eqref{eq:finiteDiff}, at the same resolution as the experimental data (i.e., $\Delta T$ = 2 K). In the limit that $\Delta T \to 0$, one recovers the exact magnetocaloric cooling rate, $\lim_{\Delta T \to 0} (\Delta M/\Delta T)_H = (\partial M/\partial T)_H$. 
A comparison between the exact and approximate magnetocaloric cooling rate obtained from mean-field is shown in Figure~\ref{fig:dSdH_MCvMFT_appendix}, for all 4 materials (computed via a converged numerical derivative with $\Delta T = 0.001$~K). 

%
%

\section{\label{ap:MC_simulations} Monte Carlo Simulations}

The classical Monte Carlo (MC) simulations have been performed on periodic lattices with $2000$--$4000$ spins. Since in the studied temperature range all materials remain in the paramagnetic state,  the simulated system sizes are sufficient to obtain the bulk behavior without additional finite-size scaling. We simulate the nearest neighbor antiferromagnetic model 
\begin{equation}
\hat{\cal H}_c = \sum_{\langle ij\rangle} \vec{s}_i\cdot \vec{s}_j 
-  \vec{H}\cdot \sum_i \vec{s}_i 
\, , 
\label{Hcl}
\end{equation}
where the spins $\vec{s}_i$ are classical vectors of unit length, $|\vec{s}_i|=1$. 
A hybrid Monte Carlo algorithm has been employed with canonical Metropolis sweeps over the lattice followed by microcanonical over-relaxation steps.
At each temperature/field point, $10^4$ Metropolis steps were used for equilibration with subsequent $10^5$ hybrid MC steps for measurements. In addition, we performed averaging over $10^2$ independent MC runs. Relative statistical errors do not exceed $0.5$\% for all obtained quantities. 

The approximate magnetocaloric cooling rate $(\Delta M/\Delta T)_H$ was computed from the magnetization using the finite difference method, Equation~\eqref{eq:finiteDiff}, at the resolution of the experimental data (i.e., temperature steps of $2$~K). 
The exact magnetocaloric cooling rate $(\partial M/\partial T)_H$ was computed from the energy-magnetization cumulant:
\begin{equation}
\biggl(\frac{\partial S_m}{\partial H}\biggr)_T = \biggl(\frac{\partial M}{\partial T}\biggr)_H =
\frac{1}{T^2} \left(\vphantom{\sum}
\langle EM\rangle -  \langle E\rangle \langle M\rangle 
\right) \, ,
\end{equation} 
where $\langle \ldots\rangle$ denotes statistical averaging. 
%
%
Figure~\ref{fig:dSdH_MCvMFT_appendix} shows a comparison between the exact and approximate values, for reference. 
The figure also allows for a comparison between the classical Monte Carlo results and the mean-field modelling in Appendix~\ref{app:meanfield}; the enhanced magnetocaloric effect in the latter is likely due to the quantised nature of the spins. 


Rescaling between dimensionless values obtained
from the classical Monte Carlo simulations~\eqref{Hcl} and real physical properties of each material was done using the conversion factors:
\begin{eqnarray}
&&\mu_0 H_{\rm exp} = \frac{J_1S}{g \mu_B}H_{\rm MC}, \ \
k_BT_{\rm exp} = J_1S(S+1) T_{\rm MC} \, , 
\nonumber \\
&& \biggl(\frac{\partial M}{\partial T}\biggr)_{\rm exp} = 
 \frac{g \mu_B N_A}{J_1(S+1)}
\biggl(\frac{\partial M}{\partial T}\biggr)_{\rm MC} 
\, . 
\end{eqnarray}
Here, $S=7/2$ and $g\approx 2$ are the spin and the $g$ factor of Gd$^{3+}$ ions, $J_1$ is the nearest neighbor exchange, $k_B$ and $N_A$ are the Boltzmann constant and Avogadro's number.
%
%

\section{\label{app:methods}Experimental Methods}
%
%

\subsection{Sample preparation}

Polycrystalline samples were prepared using a solid-state method consistent with prior reports in the literature~\cite{Sackville_Hamilton_2014,Koskelo_2022,Paddison_Gd2Sn2O7_2017,Dun_kagome_2017}. Reactants oxides were ground using a mortar and pestle and then heated in air in an alumina crucible at $T \sim 1300$\textdegree~C for several days with intermittent grindings to ensure a complete reaction. As described in Appendix~\ref{appendix:Gd_structures_soft_modes}, x-ray diffraction indicates phase pure samples ($< 1$~weight~\% impurities of Gd$_3$SbO$_7$ in Ba$_2$GdSbO$_6$) with crystal structures consistent with the literature reported values.
%
%

\subsection{X-ray diffraction and crystal structure refinements}

Room temperature powder x-ray diffraction (XRD) measurements were conducted using a Bruker D8 Advance diffractometer (Cu K$\alpha$ radiation, $\lambda = 1.54$~\AA). Data was collected with $d(2\theta)=0.01$\textdegree \hspace{1pt} from $2\theta=15-150$\textdegree, with an overall collection time of 2-3 hours. Rietveld refinements~\cite{McCusker_RietveldRefinement} of the powder XRD data were completed using the Diffrac.Suite TOPAS5 program~\cite{TOPAS_Academic}. The background was modeled using a 13-term Chebyshev polynomial and peak shapes were fit using a pseudo-Voigt function~\cite{TCHZ_peaks}. All Debye-Waller factors were set constant to the literature reported values.
%
%

\subsection{Bulk magnetic measurements}

Bulk magnetic measurements of the magnetic susceptibility $\chi(T) = dM/dH$ were conducted using a Quantum Design Magnetic Properties Measurement System (MPMS) with a superconducting interference device (SQUID). Susceptibility measurements were made at $\mu_0 H=0.01$~T, in the low-field limit where $\chi(T) = dM/dH \approx M/H$ in zero-field-cooled conditions from $1.8$~K to $300$~K. $M(H)$ measurements (described below) are linear at this field, confirming that this linear approximation of $\chi$ is valid. Isothermal magnetization $M(H)$ was measured using the ACMS option of a Quantum Design Physical Properties Measurement System (PPMS) for Gd$_3$Mg$_2$Sb$_3$O$_{14}$, Gd$_3$Ga$_5$O$_{12}$, and Gd$_2$Sn$_2$O$_7$ and a Quantum Design MPMS SQUID for Ba$_2$GdSbO$_6$. $M(H)$ measurements were made at temperatures of $2$~K to $22$~K in steps of $2$~K over a field range of $0$~T to $9$~T in steps of $0.2$~T (only up to $7$~T for Ba$_2$GdSbO$_6$). 
%
%

\subsection{Magnetocaloric effect calculations}

The magnetic entropy change for a field $H_{\rm max}$ relative to zero field was calculated from the measured $M(H)$ by first computing the temperature derivative of the magnetization using the finite differences approximation:
\begin{equation}
   \left. \left( \frac{\partial M(T,H)}{\partial T} \right)_H \right \vert_{T = T_i} \approx \frac{M(T_{i+1},H)-M(T_i,H)}{T_{i+1} - T_i} \, ,
   \label{eq:finiteDiff}
\end{equation}
and then integrating across fields as in Equation~\eqref{eq:deltaSm_fromdMdT}. The magnetization data $M(H)$ were linearly interpolated along the field direction in steps of $0.1$~T prior to extracting the magnetic entropy. 
%

\section{\label{app:entropypervol} Magnetocaloric performance per mass and volume}

\begin{table*}[htbp]
    \centering
\caption{Measured magnetocaloric effect at T=2 K and $\mu_0 H$=9 T for the corner-sharing geometries, and $\mu_0 H$=7 T for the \emph{fcc} lattice, normalized by volume and by mass. Mass densities $\rho$ were determined from Rietveld refinement for the kagome, garnet, and \emph{fcc} lattices (see Appendix \ref{appendix:Gd_structures_soft_modes}) and from the literature reported value for the pyrochlore \cite{kennedy_structural_1997}.}
    \begin{ruledtabular}
    \begin{tabular}{l|c|c|c|c|c}
     Compound & $\Delta S_{m}$ (J K$^{-1}$ mol$_{\textrm{Gd}}^{-1}$) & Molar Mass (g mol$^{-1}$) & $\rho$ (g cm$^{-3}$) & $\Delta S_{m}$ (mJ K$^{-1}$ cm$^{-3}$) & $\Delta S_{m}$ (mJ K$^{-1}$ kg$^{-1}$) \\
     \hline
     Gd$_3$Mg$_2$Sb$_3$O$_{14}$ & $0.71 R\ln(2S+1)$ & 1109.63 & 6.75 & 220 & 33 \\
     Gd$_2$Sn$_2$O$_7$ &  $0.62 R\ln(2S+1)$ & 663.92 & 7.72 & 250 & 32 \\
     Gd$_3$Ga$_5$O$_{12}$ & $0.77 R\ln(2S+1)$ & 1012.36 & 7.09 & 280 & 39 \\
     Ba$_2$GdSbO$_6$ &   $0.92 R\ln(2S+1)$ & 649.66  & 7.09 & 170 & 24 \\
    \end{tabular}
    \end{ruledtabular}
    \label{table:entropy-per-vol}
\end{table*}

The measured magnetocaloric effect, Table \ref{table:entropy-per-vol}, at T= 2 K for a field change of 9 T to 0 T for Gd$_3$Mg$_2$Sb$_3$O$_{14}$ (kagome), Gd$_2$Sn$_2$O$_7$ (pyrochlore), and Gd$_3$Ga$_5$O$_{12}$ (garnet), and for field change of 7 T to 0 T for Ba$_2$GdSbO$_6$ (\emph{fcc}), normalized by volume and by mass.


\section{\label{appendix:Gd_structures_soft_modes}
Structural refinements of Gd compounds}

\begin{figure}[htbp]
    \centering
    \includegraphics[width = 0.8\columnwidth]{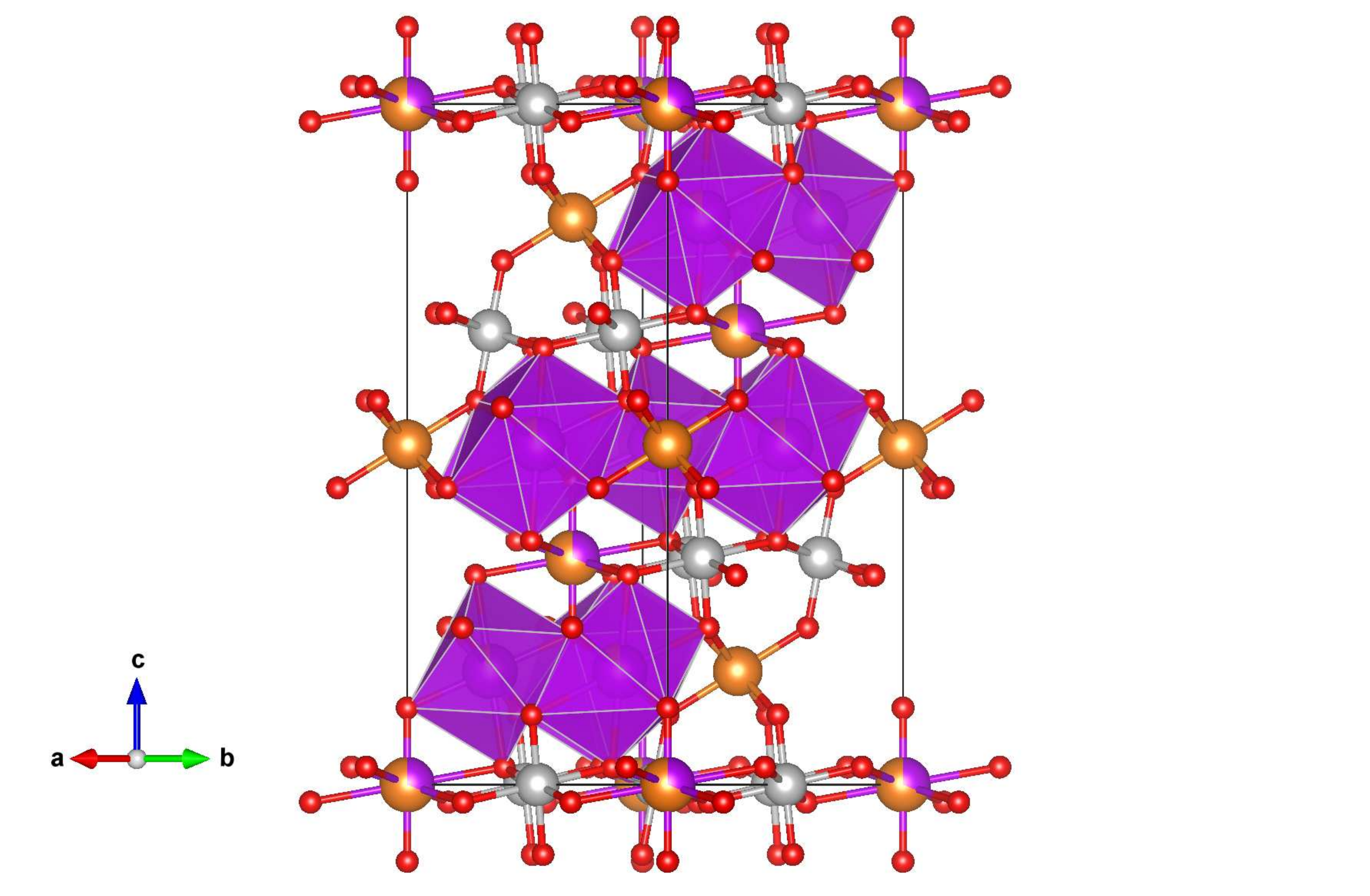}
    \caption{Refined crystal structure of Gd$_3$Mg$_2$Sb$_3$O$_{14}$ sample. The compound exhibits some site disorder, with 10.5(2)\% of Mg$^{2+}$ ions from the 3$a$ site lying on the kagome Gd$^{3+}$ $9d$ site, Table~\ref{table:GMSO_Rietveld}. Gd$^{3+}$ ions and corresponding O-polyhedra are shown in purple, Mg$^{2+}$ ions in orange, Sb$^{5+}$ ions in gray, and O$^{2-}$ ions in red. This figure was generated from the refined crystal structure file using VESTA \cite{momma_vesta_2011}. }
    \label{fig:crystal_GMSO}
\end{figure}

Rietveld analysis of the XRD shows that the refined structures are consistent with those reported in the literature~\cite{Sackville_Hamilton_2014,Karunadasa_PNAS,Paddison_Gd2Sn2O7_2017,Wellm_GMSO_2020,Dun_kagome_2017}. Recently, the family of kagome compounds $Ln_3$Mg$_2$Sb$_3$O$_{14}$ has been found to exhibit sample-dependent cation site disorder, in which the $Ln^{3+}$ ion on the 9$d$ Wyckoff site of a kagome layer swaps places with a Mg$^{2+}$ ion on the interlayer $3a$ site (see Figure~\ref{fig:crystal_GMSO})~\cite{DMSO_monopoles_2016,HMSO_2020_disorder}. We find a 10.5(2)\% cation site-disordering in the Gd$_3$Mg$_2$Sb$_3$O$_{14}$ sample in this report. A comparable amount of site disorder has been found in other samples of $Ln_3$Mg$_2$Sb$_3$O$_{14}$ including the emergent-charge-ordered kagome Ising magnet Dy$_3$Mg$_2$Sb$_3$O$_{14}$~\cite{DMSO_monopoles_2016} and the dipolar kagome ice Ho$_3$Mg$_2$Sb$_3$O$_{14}$~\cite{HMSO_2020_disorder}. 

\begin{table}[htbp]
\caption{Structural parameters of Gd$_3$Mg$_2$Sb$_3$O$_{14}$. 
PXRD refinements were carried out in the space group $R\bar{3}m$, with Mg(1) on the 
$3a$ sites $(0,0,0)$, 
Mg(2) on the $3b$ sites $(0,0,\sfrac{1}{2})$, 
Gd(1) on the 9$d$ sites $(\sfrac{1}{2},0,\sfrac{1}{2})$, 
Sb on the $9e$ sites $(\sfrac{1}{2},0,0)$, and 
O(1) on the 6$c$ sites $(0,0,z)$, and O(2) and O(3) on the 18$h$ sites ($x$,$\bar{x}$,$z$). 
Partial occupancy between the 3a and 9d sites was also refined according to the formula
[Gd(1)$_{1-x}$Mg(3)$_x$]$_{9d}$[Mg(1)$_{1-3x}$Gd(2)$_{3x}$]$_{3a}$ as in Reference~\onlinecite{DMSO_monopoles_2016}. }
\centering
\label{table:GMSO_Rietveld}
\begin{ruledtabular}
\begin{tabular}{c c c}
Gd$_3$Mg$_2$Sb$_3$O$_{14}$ & $R\bar{3}m$ & \\
\hline
$a$ (\AA) & & 7.3634(1) \\
$c$ (\AA) & & 17.4511(3) \\
Mg(1)/Gd(2) (3a) & Occupancy & 0.686(6), 0.314(6) \\
Gd(1)/Mg(3) (9d) & Occupancy & 0.895(2), 0.105(2) \\
O(1) & $z$ & 0.113(1) \\
O(2) & $x$ & 0.5261(7) \\
& $z$ & 0.8974(6) \\
O(3) & $x$ & 0.4714(8) \\
& $z$ & 0.3591(5) \\
Overall $B_{iso}$ (\AA$^2$) & & 0.76(2) \\
Gd$_3$SbO$_7$ (wt \%) & & 0.83(4) \\
$R_{wp}$ & & 9.69 \\
$\chi^2$ & & 1.42 \\
\end{tabular}
\end{ruledtabular}
\end{table}       

\begin{figure}[htbp]
    \centering
    \includegraphics[width=\columnwidth]{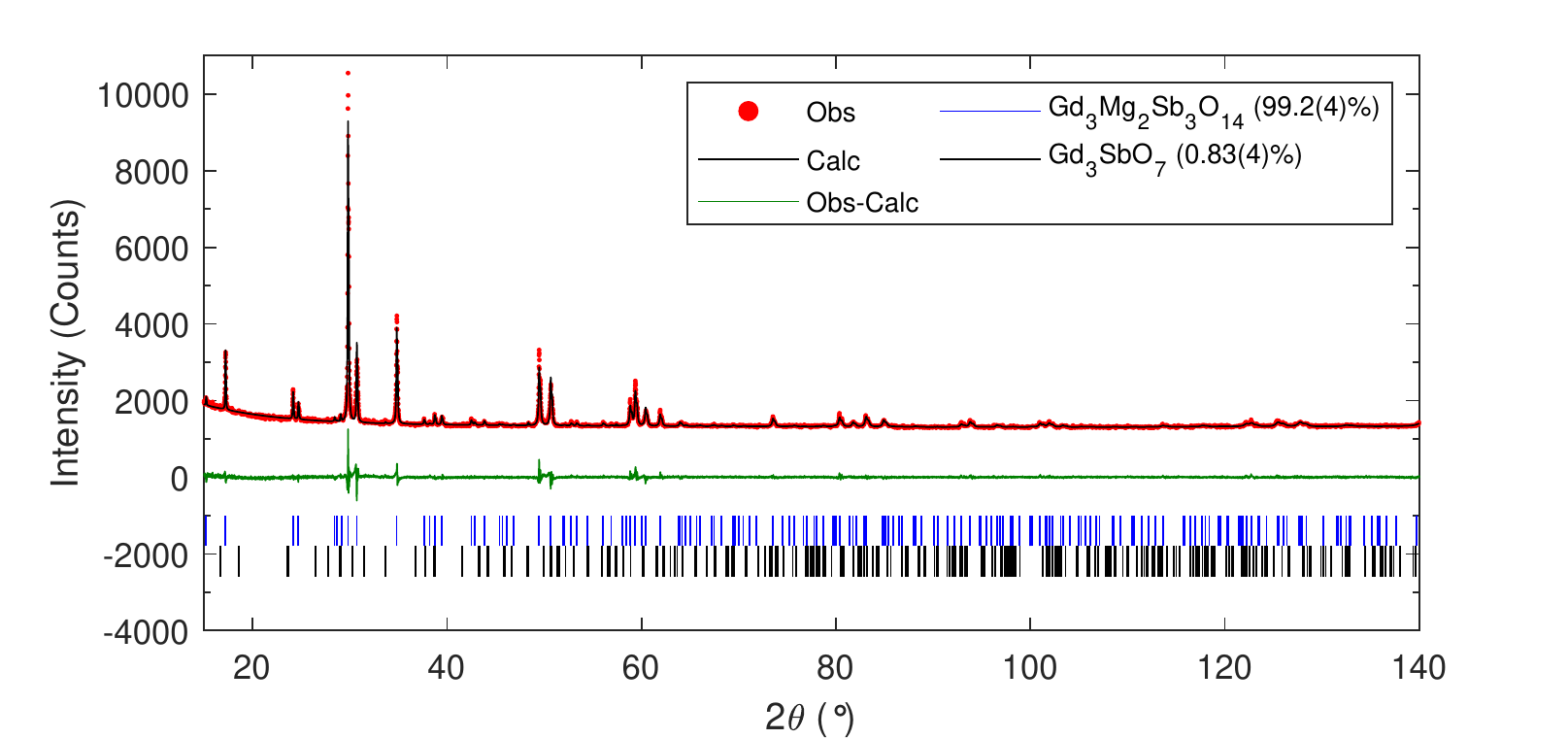}
   \caption{Rietveld refinement of Gd$_3$Mg$_2$Sb$_3$O$_{14}$ from PXRD.}
    \label{fig:XRD_GMSO}
\end{figure}

\begin{figure}[htbp]
    \centering
\includegraphics[width=\columnwidth]{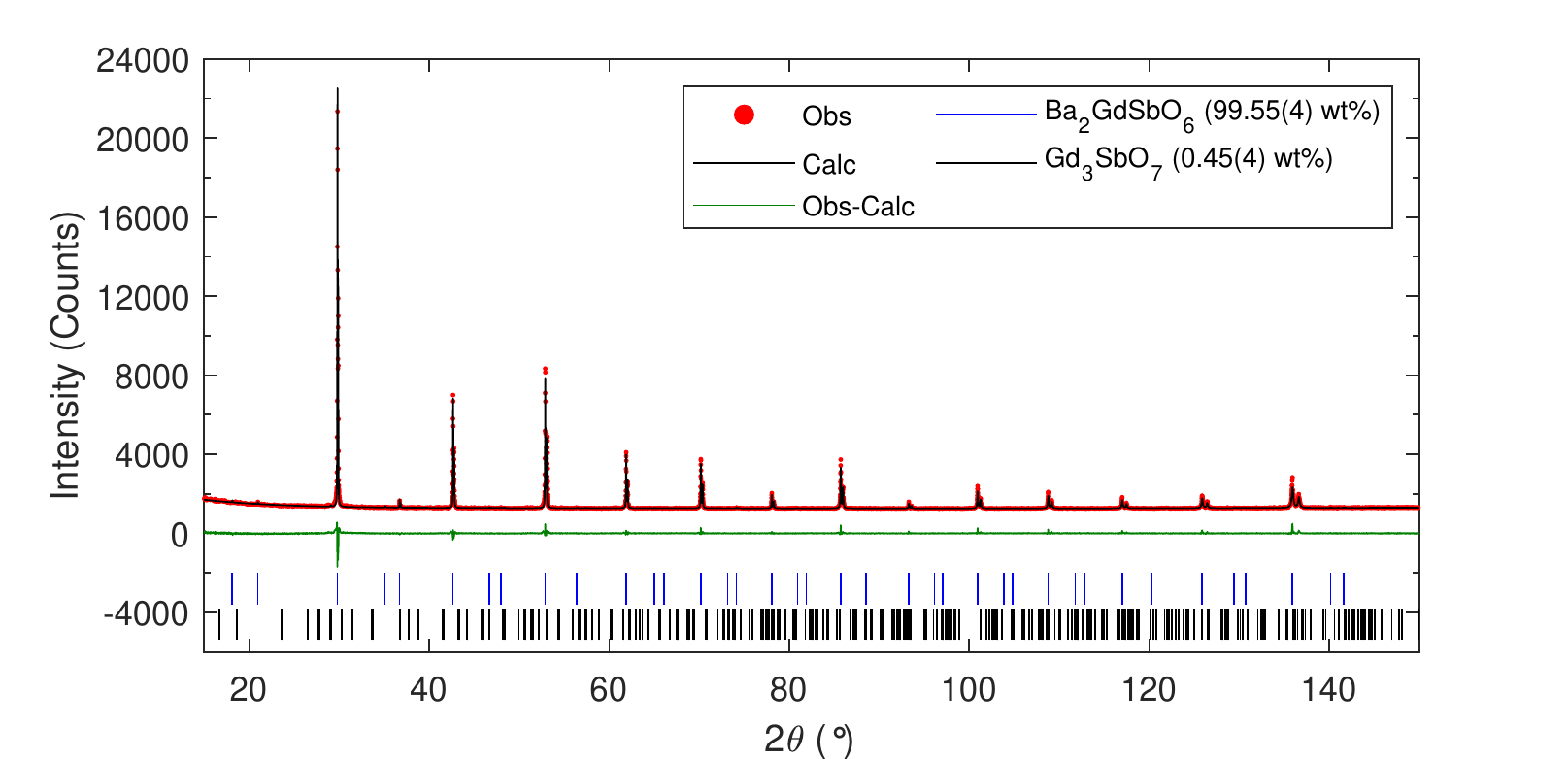}
   \caption{Rietveld refinement of Ba$_2$GdSbO$_6$ from PXRD.}
    \label{fig:XRD_BaGd}
\end{figure}

\begin{figure}[htbp]
    \centering   \includegraphics[width=\columnwidth]{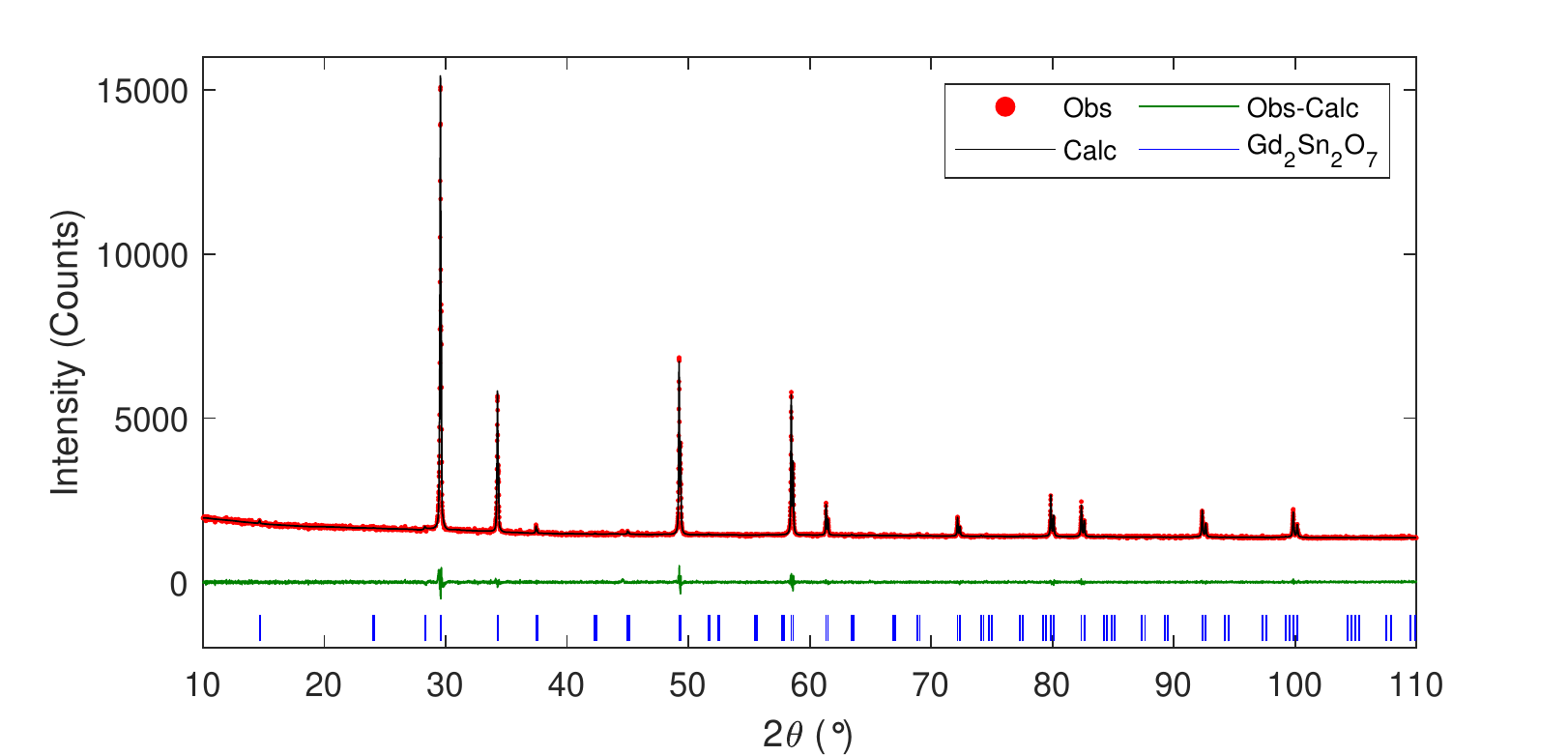}
   \caption{Rietveld refinement of Gd$_2$Sn$_2$O$_7$ from PXRD.}
    \label{fig:XRD_GdSnO}
\end{figure}

\begin{figure}[htbp]
    \centering
\includegraphics[width=\columnwidth]{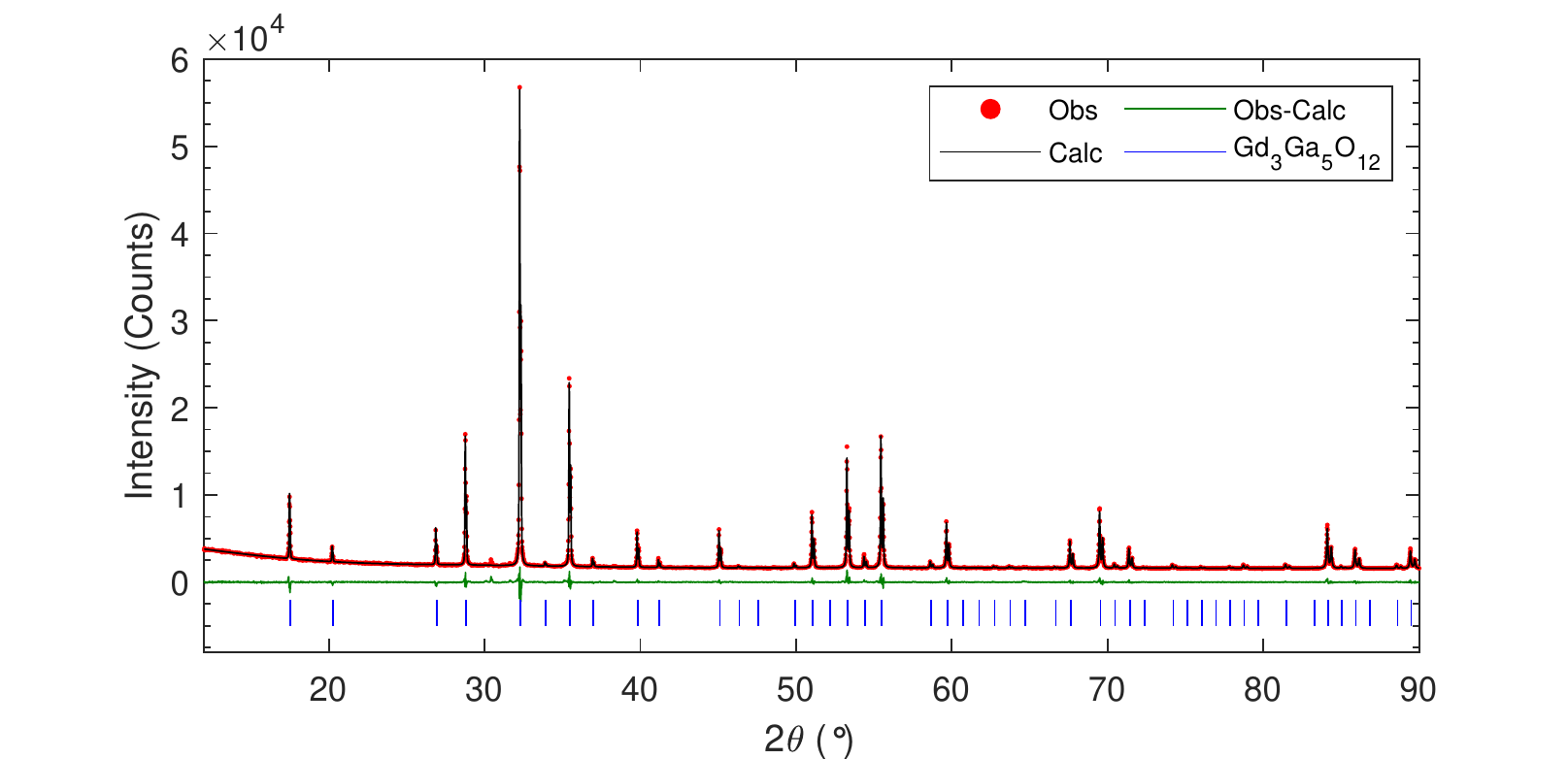}
   \caption{Rietveld refinement of Gd$_3$Ga$_5$O$_{12}$ from PXRD.}
    \label{fig:XRD_GGG}
\end{figure}

\newpage

\bibliography{bibfile}

\begin{thebibliography}{44}%
\makeatletter
\providecommand \@ifxundefined [1]{%
 \@ifx{#1\undefined}
}%
\providecommand \@ifnum [1]{%
 \ifnum #1\expandafter \@firstoftwo
 \else \expandafter \@secondoftwo
 \fi
}%
\providecommand \@ifx [1]{%
 \ifx #1\expandafter \@firstoftwo
 \else \expandafter \@secondoftwo
 \fi
}%
\providecommand \natexlab [1]{#1}%
\providecommand \enquote  [1]{``#1''}%
\providecommand \bibnamefont  [1]{#1}%
\providecommand \bibfnamefont [1]{#1}%
\providecommand \citenamefont [1]{#1}%
\providecommand \href@noop [0]{\@secondoftwo}%
\providecommand \href [0]{\begingroup \@sanitize@url \@href}%
\providecommand \@href[1]{\@@startlink{#1}\@@href}%
\providecommand \@@href[1]{\endgroup#1\@@endlink}%
\providecommand \@sanitize@url [0]{\catcode `\\12\catcode `\$12\catcode
  `\&12\catcode `\#12\catcode `\^12\catcode `\_12\catcode `\%12\relax}%
\providecommand \@@startlink[1]{}%
\providecommand \@@endlink[0]{}%
\providecommand \url  [0]{\begingroup\@sanitize@url \@url }%
\providecommand \@url [1]{\endgroup\@href {#1}{\urlprefix }}%
\providecommand \urlprefix  [0]{URL }%
\providecommand \Eprint [0]{\href }%
\providecommand \doibase [0]{https://doi.org/}%
\providecommand \selectlanguage [0]{\@gobble}%
\providecommand \bibinfo  [0]{\@secondoftwo}%
\providecommand \bibfield  [0]{\@secondoftwo}%
\providecommand \translation [1]{[#1]}%
\providecommand \BibitemOpen [0]{}%
\providecommand \bibitemStop [0]{}%
\providecommand \bibitemNoStop [0]{.\EOS\space}%
\providecommand \EOS [0]{\spacefactor3000\relax}%
\providecommand \BibitemShut  [1]{\csname bibitem#1\endcsname}%
\let\auto@bib@innerbib\@empty
\bibitem [{\citenamefont {Moya}\ and\ \citenamefont
  {Mathur}(2020)}]{Science_Moya_Caloric}%
  \BibitemOpen
  \bibfield  {author} {\bibinfo {author} {\bibfnamefont {X.}~\bibnamefont
  {Moya}}\ and\ \bibinfo {author} {\bibfnamefont {N.~D.}\ \bibnamefont
  {Mathur}},\ }\bibfield  {title} {\bibinfo {title} {Caloric materials for
  cooling and heating},\ }\href {https://doi.org/10.1126/science.abb0973}
  {\bibfield  {journal} {\bibinfo  {journal} {Science}\ }\textbf {\bibinfo
  {volume} {370}},\ \bibinfo {pages} {797} (\bibinfo {year}
  {2020})}\BibitemShut {NoStop}%
\bibitem [{\citenamefont {Wikus}\ \emph {et~al.}(2014)\citenamefont {Wikus},
  \citenamefont {Canavan}, \citenamefont {Heine}, \citenamefont {Matsumoto},\
  and\ \citenamefont {Numazawa}}]{MCE_review_PMsalts}%
  \BibitemOpen
  \bibfield  {author} {\bibinfo {author} {\bibfnamefont {P.}~\bibnamefont
  {Wikus}}, \bibinfo {author} {\bibfnamefont {E.}~\bibnamefont {Canavan}},
  \bibinfo {author} {\bibfnamefont {S.~T.}\ \bibnamefont {Heine}}, \bibinfo
  {author} {\bibfnamefont {K.}~\bibnamefont {Matsumoto}},\ and\ \bibinfo
  {author} {\bibfnamefont {T.}~\bibnamefont {Numazawa}},\ }\bibfield  {title}
  {\bibinfo {title} {Magnetocaloric materials and the optimization of cooling
  power density},\ }\href
  {https://doi.org/https://doi.org/10.1016/j.cryogenics.2014.04.005} {\bibfield
   {journal} {\bibinfo  {journal} {Cryogenics}\ }\textbf {\bibinfo {volume}
  {62}},\ \bibinfo {pages} {150} (\bibinfo {year} {2014})}\BibitemShut
  {NoStop}%
\bibitem [{\citenamefont {Zhitomirsky}(2003)}]{Zhitomirsky_2003}%
  \BibitemOpen
  \bibfield  {author} {\bibinfo {author} {\bibfnamefont {M.}~\bibnamefont
  {Zhitomirsky}},\ }\bibfield  {title} {\bibinfo {title} {Enhanced
  magnetocaloric effect in frustrated magnets},\ }\href
  {https://doi.org/10.1103/PhysRevB.67.104421} {\bibfield  {journal} {\bibinfo
  {journal} {Phys. Rev. B}\ }\textbf {\bibinfo {volume} {67}},\ \bibinfo
  {pages} {104421} (\bibinfo {year} {2003})}\BibitemShut {NoStop}%
\bibitem [{\citenamefont {Chen}\ \emph {et~al.}(2015)\citenamefont {Chen},
  \citenamefont {Prokleška}, \citenamefont {Xu}, \citenamefont {Liu},
  \citenamefont {Liu}, \citenamefont {Zhang}, \citenamefont {Jia},
  \citenamefont {Sechovský},\ and\ \citenamefont {Tong}}]{GdF3_MCE_2015}%
  \BibitemOpen
  \bibfield  {author} {\bibinfo {author} {\bibfnamefont {Y.-C.}\ \bibnamefont
  {Chen}}, \bibinfo {author} {\bibfnamefont {J.}~\bibnamefont {Prokleška}},
  \bibinfo {author} {\bibfnamefont {W.-J.}\ \bibnamefont {Xu}}, \bibinfo
  {author} {\bibfnamefont {J.-L.}\ \bibnamefont {Liu}}, \bibinfo {author}
  {\bibfnamefont {J.}~\bibnamefont {Liu}}, \bibinfo {author} {\bibfnamefont
  {W.-X.}\ \bibnamefont {Zhang}}, \bibinfo {author} {\bibfnamefont {J.-H.}\
  \bibnamefont {Jia}}, \bibinfo {author} {\bibfnamefont {V.}~\bibnamefont
  {Sechovský}},\ and\ \bibinfo {author} {\bibfnamefont {M.-L.}\ \bibnamefont
  {Tong}},\ }\bibfield  {title} {\bibinfo {title} {A brilliant cryogenic
  magnetic coolant: magnetic and magnetocaloric study of ferromagnetically
  coupled \ce{ GdF$_3$}},\ }\href {https://doi.org/10.1039/C5TC02352A}
  {\bibfield  {journal} {\bibinfo  {journal} {Journal of Materials Chemistry
  C}\ }\textbf {\bibinfo {volume} {3}},\ \bibinfo {pages} {12206} (\bibinfo
  {year} {2015})}\BibitemShut {NoStop}%
\bibitem [{\citenamefont {Brasiliano}(2017)}]{GGG_limits_MCE}%
  \BibitemOpen
  \bibfield  {author} {\bibinfo {author} {\bibfnamefont {D.~A.~P.}\
  \bibnamefont {Brasiliano}},\ }\emph {\bibinfo {title} {{Etude et
  r{\'e}alisation d'une ADR spatiale 4 K - 50 mK. Instrumentation et méthodes
  pour l'astrophysique.}}},\ \href
  {https://tel.archives-ouvertes.fr/tel-01688809} {\bibinfo {type} {Phd
  thesis}},\ \bibinfo  {school} {{Universit{\'e} Grenoble Alpes}} (\bibinfo
  {year} {2017})\BibitemShut {NoStop}%
\bibitem [{xu_(2022)}]{xu_gdohf2_2022}%
  \BibitemOpen
  \bibfield  {title} {\bibinfo {title} {\ce{Gd(OH)F2}: A promising cryogenic
  magnetic refrigerant},\ }\href@noop {} {\bibfield  {journal} {\bibinfo
  {journal} {Journal of the American Chemical Society}\ }\textbf {\bibinfo
  {volume} {144}},\ \bibinfo {pages} {13787} (\bibinfo {year}
  {2022})}\BibitemShut {NoStop}%
\bibitem [{\citenamefont {Lorusso}\ \emph {et~al.}(2013)\citenamefont
  {Lorusso}, \citenamefont {Sharples}, \citenamefont {Palacios}, \citenamefont
  {Roubeau}, \citenamefont {Brechin}, \citenamefont {Sessoli}, \citenamefont
  {Rossin}, \citenamefont {Tuna}, \citenamefont {McInnes}, \citenamefont
  {Collison},\ and\ \citenamefont {Evangelisti}}]{Gd-formate_2013}%
  \BibitemOpen
  \bibfield  {author} {\bibinfo {author} {\bibfnamefont {G.}~\bibnamefont
  {Lorusso}}, \bibinfo {author} {\bibfnamefont {J.}~\bibnamefont {Sharples}},
  \bibinfo {author} {\bibfnamefont {E.}~\bibnamefont {Palacios}}, \bibinfo
  {author} {\bibfnamefont {O.}~\bibnamefont {Roubeau}}, \bibinfo {author}
  {\bibfnamefont {E.}~\bibnamefont {Brechin}}, \bibinfo {author} {\bibfnamefont
  {R.}~\bibnamefont {Sessoli}}, \bibinfo {author} {\bibfnamefont
  {A.}~\bibnamefont {Rossin}}, \bibinfo {author} {\bibfnamefont
  {F.}~\bibnamefont {Tuna}}, \bibinfo {author} {\bibfnamefont {E.}~\bibnamefont
  {McInnes}}, \bibinfo {author} {\bibfnamefont {D.}~\bibnamefont {Collison}},\
  and\ \bibinfo {author} {\bibfnamefont {M.}~\bibnamefont {Evangelisti}},\
  }\bibfield  {title} {\bibinfo {title} {A dense metal–organic framework for
  enhanced magnetic refrigeration},\ }\href
  {https://doi.org/https://doi.org/10.1002/adma.201301997} {\bibfield
  {journal} {\bibinfo  {journal} {Advanced Materials}\ }\textbf {\bibinfo
  {volume} {25}},\ \bibinfo {pages} {4653} (\bibinfo {year}
  {2013})}\BibitemShut {NoStop}%
\bibitem [{\citenamefont {Palacios}\ \emph {et~al.}(2014)\citenamefont
  {Palacios}, \citenamefont {Rodr\'{\i}guez-Velamaz\'an}, \citenamefont
  {Evangelisti}, \citenamefont {McIntyre}, \citenamefont {Lorusso},
  \citenamefont {Visser}, \citenamefont {de~Jongh},\ and\ \citenamefont
  {Boatner}}]{GdPO4_MCE_2014}%
  \BibitemOpen
  \bibfield  {author} {\bibinfo {author} {\bibfnamefont {E.}~\bibnamefont
  {Palacios}}, \bibinfo {author} {\bibfnamefont {J.~A.}\ \bibnamefont
  {Rodr\'{\i}guez-Velamaz\'an}}, \bibinfo {author} {\bibfnamefont
  {M.}~\bibnamefont {Evangelisti}}, \bibinfo {author} {\bibfnamefont {G.~J.}\
  \bibnamefont {McIntyre}}, \bibinfo {author} {\bibfnamefont {G.}~\bibnamefont
  {Lorusso}}, \bibinfo {author} {\bibfnamefont {D.}~\bibnamefont {Visser}},
  \bibinfo {author} {\bibfnamefont {L.~J.}\ \bibnamefont {de~Jongh}},\ and\
  \bibinfo {author} {\bibfnamefont {L.~A.}\ \bibnamefont {Boatner}},\
  }\bibfield  {title} {\bibinfo {title} {Magnetic structure and magnetocalorics
  of \ce{GdPO4}},\ }\href {https://doi.org/10.1103/PhysRevB.90.214423}
  {\bibfield  {journal} {\bibinfo  {journal} {Phys. Rev. B}\ }\textbf {\bibinfo
  {volume} {90}},\ \bibinfo {pages} {214423} (\bibinfo {year}
  {2014})}\BibitemShut {NoStop}%
\bibitem [{\citenamefont {Reis}(2020)}]{MCE_review_2020}%
  \BibitemOpen
  \bibfield  {author} {\bibinfo {author} {\bibfnamefont {M.~S.}\ \bibnamefont
  {Reis}},\ }\bibfield  {title} {\bibinfo {title} {Magnetocaloric and
  barocaloric effects of metal complexes for solid state cooling: Review,
  trends and perspectives},\ }\href
  {https://doi.org/https://doi.org/10.1016/j.ccr.2020.213357} {\bibfield
  {journal} {\bibinfo  {journal} {Coordination Chemistry Reviews}\ }\textbf
  {\bibinfo {volume} {417}},\ \bibinfo {pages} {213357} (\bibinfo {year}
  {2020})}\BibitemShut {NoStop}%
\bibitem [{\citenamefont {Mukherjee}(2018)}]{Paromita_thesis}%
  \BibitemOpen
  \bibfield  {author} {\bibinfo {author} {\bibfnamefont {P.}~\bibnamefont
  {Mukherjee}},\ }\emph {\bibinfo {title} {Investigation of the magnetic and
  magnetocaloric properties of complex lanthanide oxides}},\ \href
  {https://www.repository.cam.ac.uk/handle/1810/275425} {Ph.D. thesis},\
  \bibinfo  {school} {University of Cambridge}, \bibinfo {address} {Cambridge,
  Cambridgeshire, UK} (\bibinfo {year} {2018}),\ \bibinfo {note} {publisher:
  Apollo - University of Cambridge Repository}\BibitemShut {NoStop}%
\bibitem [{\citenamefont {{Paixao Brasiliano}}\ \emph
  {et~al.}(2020)\citenamefont {{Paixao Brasiliano}}, \citenamefont {Duval},
  \citenamefont {Marin}, \citenamefont {Bichaud}, \citenamefont {Brison},
  \citenamefont {Zhitomirsky},\ and\ \citenamefont {Luchier}}]{YbGG_MCE}%
  \BibitemOpen
  \bibfield  {author} {\bibinfo {author} {\bibfnamefont {D.~A.}\ \bibnamefont
  {{Paixao Brasiliano}}}, \bibinfo {author} {\bibfnamefont {J.-M.}\
  \bibnamefont {Duval}}, \bibinfo {author} {\bibfnamefont {C.}~\bibnamefont
  {Marin}}, \bibinfo {author} {\bibfnamefont {E.}~\bibnamefont {Bichaud}},
  \bibinfo {author} {\bibfnamefont {J.-P.}\ \bibnamefont {Brison}}, \bibinfo
  {author} {\bibfnamefont {M.}~\bibnamefont {Zhitomirsky}},\ and\ \bibinfo
  {author} {\bibfnamefont {N.}~\bibnamefont {Luchier}},\ }\bibfield  {title}
  {\bibinfo {title} {\ce{YbGG} material for adiabatic demagnetization in the
  100 \ce{mK}–3 \ce{K} range},\ }\href
  {https://doi.org/https://doi.org/10.1016/j.cryogenics.2019.103002} {\bibfield
   {journal} {\bibinfo  {journal} {Cryogenics}\ }\textbf {\bibinfo {volume}
  {105}},\ \bibinfo {pages} {103002} (\bibinfo {year} {2020})}\BibitemShut
  {NoStop}%
\bibitem [{\citenamefont {Dun}\ \emph {et~al.}(2016)\citenamefont {Dun},
  \citenamefont {Trinh}, \citenamefont {Li}, \citenamefont {Lee}, \citenamefont
  {Chen}, \citenamefont {Baumbach}, \citenamefont {Hu}, \citenamefont {Wang},
  \citenamefont {Choi}, \citenamefont {Shastry}, \citenamefont {Ramirez},\ and\
  \citenamefont {Zhou}}]{Dun_LnMSO_2016_PRL}%
  \BibitemOpen
  \bibfield  {author} {\bibinfo {author} {\bibfnamefont {Z.~L.}\ \bibnamefont
  {Dun}}, \bibinfo {author} {\bibfnamefont {J.}~\bibnamefont {Trinh}}, \bibinfo
  {author} {\bibfnamefont {K.}~\bibnamefont {Li}}, \bibinfo {author}
  {\bibfnamefont {M.}~\bibnamefont {Lee}}, \bibinfo {author} {\bibfnamefont
  {K.~W.}\ \bibnamefont {Chen}}, \bibinfo {author} {\bibfnamefont
  {R.}~\bibnamefont {Baumbach}}, \bibinfo {author} {\bibfnamefont {Y.~F.}\
  \bibnamefont {Hu}}, \bibinfo {author} {\bibfnamefont {Y.~X.}\ \bibnamefont
  {Wang}}, \bibinfo {author} {\bibfnamefont {E.~S.}\ \bibnamefont {Choi}},
  \bibinfo {author} {\bibfnamefont {B.~S.}\ \bibnamefont {Shastry}}, \bibinfo
  {author} {\bibfnamefont {A.~P.}\ \bibnamefont {Ramirez}},\ and\ \bibinfo
  {author} {\bibfnamefont {H.~D.}\ \bibnamefont {Zhou}},\ }\bibfield  {title}
  {\bibinfo {title} {Magnetic ground states of the rare-earth tripod kagome
  lattice \ce{Mg2RESb3O_{14}}},\ }\href
  {https://doi.org/10.1103/PhysRevLett.116.157201} {\bibfield  {journal}
  {\bibinfo  {journal} {Phys. Rev. Lett.}\ }\textbf {\bibinfo {volume} {116}},\
  \bibinfo {pages} {157201} (\bibinfo {year} {2016})}\BibitemShut {NoStop}%
\bibitem [{\citenamefont {Petrenko}\ and\ \citenamefont
  {McK.~Paul}(2000)}]{Petrenko_GGG_Dipolar_term_2000}%
  \BibitemOpen
  \bibfield  {author} {\bibinfo {author} {\bibfnamefont {O.~A.}\ \bibnamefont
  {Petrenko}}\ and\ \bibinfo {author} {\bibfnamefont {D.}~\bibnamefont
  {McK.~Paul}},\ }\bibfield  {title} {\bibinfo {title} {Classical heisenberg
  antiferromagnet on a garnet lattice: A monte carlo simulation},\ }\href@noop
  {} {\bibfield  {journal} {\bibinfo  {journal} {Phys. Rev. B}\ }\textbf
  {\bibinfo {volume} {63}},\ \bibinfo {pages} {024409} (\bibinfo {year}
  {2000})}\BibitemShut {NoStop}%
\bibitem [{\citenamefont {Koskelo}\ \emph {et~al.}(2022)\citenamefont
  {Koskelo}, \citenamefont {Liu}, \citenamefont {Mukherjee}, \citenamefont
  {Kelly},\ and\ \citenamefont {Dutton}}]{Koskelo_2022}%
  \BibitemOpen
  \bibfield  {author} {\bibinfo {author} {\bibfnamefont {E.~C.}\ \bibnamefont
  {Koskelo}}, \bibinfo {author} {\bibfnamefont {C.}~\bibnamefont {Liu}},
  \bibinfo {author} {\bibfnamefont {P.}~\bibnamefont {Mukherjee}}, \bibinfo
  {author} {\bibfnamefont {N.~D.}\ \bibnamefont {Kelly}},\ and\ \bibinfo
  {author} {\bibfnamefont {S.~E.}\ \bibnamefont {Dutton}},\ }\bibfield  {title}
  {\bibinfo {title} {Free-spin dominated magnetocaloric effect in dense
  \ce{Gd${^3+}$} double perovskites},\ }\href
  {https://doi.org/10.1021/acs.chemmater.2c00261} {\bibfield  {journal}
  {\bibinfo  {journal} {Chemistry of Materials}\ }\textbf {\bibinfo {volume}
  {34}},\ \bibinfo {pages} {3440} (\bibinfo {year} {2022})}\BibitemShut
  {NoStop}%
\bibitem [{\citenamefont {Zhitomirsky}(2015)}]{zhitomirsky2015real}%
  \BibitemOpen
  \bibfield  {author} {\bibinfo {author} {\bibfnamefont {M.}~\bibnamefont
  {Zhitomirsky}},\ }\bibfield  {title} {\bibinfo {title} {Real-space
  perturbation theory for frustrated magnets: application to magnetization
  plateaus},\ }in\ \href@noop {} {\emph {\bibinfo {booktitle} {Journal of
  Physics: Conference Series}}},\ Vol.\ \bibinfo {volume} {592}\ (\bibinfo
  {organization} {IOP Publishing},\ \bibinfo {year} {2015})\ p.\ \bibinfo
  {pages} {012110}\BibitemShut {NoStop}%
\bibitem [{\citenamefont {Jackeli}\ and\ \citenamefont
  {Zhitomirsky}(2004)}]{fcc_magnons_atHsat_Zhitomirksy}%
  \BibitemOpen
  \bibfield  {author} {\bibinfo {author} {\bibfnamefont {G.}~\bibnamefont
  {Jackeli}}\ and\ \bibinfo {author} {\bibfnamefont {M.~E.}\ \bibnamefont
  {Zhitomirsky}},\ }\bibfield  {title} {\bibinfo {title} {Frustrated
  {Antiferromagnets} at {High} {Fields}: {Bose}-{Einstein} {Condensation} in
  {Degenerate} {Spectra}},\ }\href
  {https://doi.org/10.1103/PhysRevLett.93.017201} {\bibfield  {journal}
  {\bibinfo  {journal} {Physical Review Letters}\ }\textbf {\bibinfo {volume}
  {93}},\ \bibinfo {pages} {017201} (\bibinfo {year} {2004})}\BibitemShut
  {NoStop}%
\bibitem [{\citenamefont {Sosin}\ \emph
  {et~al.}(2005{\natexlab{a}})\citenamefont {Sosin}, \citenamefont {Prozorova},
  \citenamefont {Smirnov}, \citenamefont {Golov}, \citenamefont {Berkutov},
  \citenamefont {Petrenko}, \citenamefont {Balakrishnan},\ and\ \citenamefont
  {Zhitomirsky}}]{Gd2Ti2O7_MCE_Tad_Zhitomirksy}%
  \BibitemOpen
  \bibfield  {author} {\bibinfo {author} {\bibfnamefont {S.~S.}\ \bibnamefont
  {Sosin}}, \bibinfo {author} {\bibfnamefont {L.~A.}\ \bibnamefont
  {Prozorova}}, \bibinfo {author} {\bibfnamefont {A.~I.}\ \bibnamefont
  {Smirnov}}, \bibinfo {author} {\bibfnamefont {A.~I.}\ \bibnamefont {Golov}},
  \bibinfo {author} {\bibfnamefont {I.~B.}\ \bibnamefont {Berkutov}}, \bibinfo
  {author} {\bibfnamefont {O.~A.}\ \bibnamefont {Petrenko}}, \bibinfo {author}
  {\bibfnamefont {G.}~\bibnamefont {Balakrishnan}},\ and\ \bibinfo {author}
  {\bibfnamefont {M.~E.}\ \bibnamefont {Zhitomirsky}},\ }\bibfield  {title}
  {\bibinfo {title} {Magnetocaloric effect in pyrochlore antiferromagnet
  \ce{Gd2Ti2O7}},\ }\href {https://doi.org/10.1103/PhysRevB.71.094413}
  {\bibfield  {journal} {\bibinfo  {journal} {Phys. Rev. B}\ }\textbf {\bibinfo
  {volume} {71}},\ \bibinfo {pages} {094413} (\bibinfo {year}
  {2005}{\natexlab{a}})}\BibitemShut {NoStop}%
\bibitem [{\citenamefont {Zhitomirsky}\ and\ \citenamefont
  {Tsunetsugu}(2005)}]{zhitomirsky_highFields_2005}%
  \BibitemOpen
  \bibfield  {author} {\bibinfo {author} {\bibfnamefont {M.}~\bibnamefont
  {Zhitomirsky}}\ and\ \bibinfo {author} {\bibfnamefont {H.}~\bibnamefont
  {Tsunetsugu}},\ }\bibfield  {title} {\bibinfo {title} {High {Field}
  {Properties} of {Geometrically} {Frustrated} {Magnets}},\ }\href
  {https://doi.org/10.1143/PTPS.160.361} {\bibfield  {journal} {\bibinfo
  {journal} {Progress of Theoretical Physics Supplement}\ }\textbf {\bibinfo
  {volume} {160}},\ \bibinfo {pages} {361} (\bibinfo {year}
  {2005})}\BibitemShut {NoStop}%
\bibitem [{\citenamefont {Schick}\ \emph {et~al.}(2020)\citenamefont {Schick},
  \citenamefont {Ziman},\ and\ \citenamefont
  {Zhitomirsky}}]{fcc_orderbydisorder_2020}%
  \BibitemOpen
  \bibfield  {author} {\bibinfo {author} {\bibfnamefont {R.}~\bibnamefont
  {Schick}}, \bibinfo {author} {\bibfnamefont {T.}~\bibnamefont {Ziman}},\ and\
  \bibinfo {author} {\bibfnamefont {M.~E.}\ \bibnamefont {Zhitomirsky}},\
  }\bibfield  {title} {\bibinfo {title} {Quantum versus thermal fluctuations in
  the fcc antiferromagnet: {Alternative} routes to order by disorder},\ }\href
  {https://doi.org/10.1103/PhysRevB.102.220405} {\bibfield  {journal} {\bibinfo
   {journal} {Physical Review B}\ }\textbf {\bibinfo {volume} {102}},\ \bibinfo
  {pages} {220405} (\bibinfo {year} {2020})}\BibitemShut {NoStop}%
\bibitem [{\citenamefont {Dun}\ \emph {et~al.}(2017)\citenamefont {Dun},
  \citenamefont {Trinh}, \citenamefont {Lee}, \citenamefont {Choi},
  \citenamefont {Li}, \citenamefont {Hu}, \citenamefont {Wang}, \citenamefont
  {Blanc}, \citenamefont {Ramirez},\ and\ \citenamefont
  {Zhou}}]{Dun_kagome_2017}%
  \BibitemOpen
  \bibfield  {author} {\bibinfo {author} {\bibfnamefont {Z.~L.}\ \bibnamefont
  {Dun}}, \bibinfo {author} {\bibfnamefont {J.}~\bibnamefont {Trinh}}, \bibinfo
  {author} {\bibfnamefont {M.}~\bibnamefont {Lee}}, \bibinfo {author}
  {\bibfnamefont {E.~S.}\ \bibnamefont {Choi}}, \bibinfo {author}
  {\bibfnamefont {K.}~\bibnamefont {Li}}, \bibinfo {author} {\bibfnamefont
  {Y.~F.}\ \bibnamefont {Hu}}, \bibinfo {author} {\bibfnamefont {Y.~X.}\
  \bibnamefont {Wang}}, \bibinfo {author} {\bibfnamefont {N.}~\bibnamefont
  {Blanc}}, \bibinfo {author} {\bibfnamefont {A.~P.}\ \bibnamefont {Ramirez}},\
  and\ \bibinfo {author} {\bibfnamefont {H.~D.}\ \bibnamefont {Zhou}},\
  }\bibfield  {title} {\bibinfo {title} {Structural and magnetic properties of
  two branches of the tripod-kagome-lattice family \ce{A2R3Sb3O_{{14}}} ({A =
  Mg, Zn; R = Pr, Nd, Gd, Tb, Dy, Ho, Er, Yb})},\ }\href
  {https://doi.org/10.1103/PhysRevB.95.104439} {\bibfield  {journal} {\bibinfo
  {journal} {Phys. Rev. B}\ }\textbf {\bibinfo {volume} {95}},\ \bibinfo
  {pages} {104439} (\bibinfo {year} {2017})}\BibitemShut {NoStop}%
\bibitem [{\citenamefont {Wellm}\ \emph {et~al.}(2020)\citenamefont {Wellm},
  \citenamefont {Zeisner}, \citenamefont {Alfonsov}, \citenamefont {Sturza},
  \citenamefont {Bastien}, \citenamefont {Ga\ss{}}, \citenamefont {Wurmehl},
  \citenamefont {Wolter}, \citenamefont {B\"uchner},\ and\ \citenamefont
  {Kataev}}]{Wellm_GMSO_2020}%
  \BibitemOpen
  \bibfield  {author} {\bibinfo {author} {\bibfnamefont {C.}~\bibnamefont
  {Wellm}}, \bibinfo {author} {\bibfnamefont {J.}~\bibnamefont {Zeisner}},
  \bibinfo {author} {\bibfnamefont {A.}~\bibnamefont {Alfonsov}}, \bibinfo
  {author} {\bibfnamefont {M.-I.}\ \bibnamefont {Sturza}}, \bibinfo {author}
  {\bibfnamefont {G.}~\bibnamefont {Bastien}}, \bibinfo {author} {\bibfnamefont
  {S.}~\bibnamefont {Ga\ss{}}}, \bibinfo {author} {\bibfnamefont
  {S.}~\bibnamefont {Wurmehl}}, \bibinfo {author} {\bibfnamefont {A.~U.~B.}\
  \bibnamefont {Wolter}}, \bibinfo {author} {\bibfnamefont {B.}~\bibnamefont
  {B\"uchner}},\ and\ \bibinfo {author} {\bibfnamefont {V.}~\bibnamefont
  {Kataev}},\ }\bibfield  {title} {\bibinfo {title} {Magnetic interactions in
  the tripod kagome antiferromagnet \ce{Mg2Gd3Sb3O_{14}} probed by static
  magnetometry and high-field \ce{ESR} spectroscopy},\ }\href@noop {}
  {\bibfield  {journal} {\bibinfo  {journal} {Phys. Rev. B}\ }\textbf {\bibinfo
  {volume} {102}},\ \bibinfo {pages} {214414} (\bibinfo {year}
  {2020})}\BibitemShut {NoStop}%
\bibitem [{\citenamefont {Bondah-Jagalu}\ and\ \citenamefont
  {Bramwell}(2001)}]{bondah-jagalu_magnetic_2001}%
  \BibitemOpen
  \bibfield  {author} {\bibinfo {author} {\bibfnamefont {V.}~\bibnamefont
  {Bondah-Jagalu}}\ and\ \bibinfo {author} {\bibfnamefont {S.~T.}\ \bibnamefont
  {Bramwell}},\ }\bibfield  {title} {\bibinfo {title} {Magnetic susceptibility
  study of the heavy rare-earth stannate pyrochlores},\ }\href
  {https://doi.org/10.1139/p01-118} {\bibfield  {journal} {\bibinfo  {journal}
  {Canadian Journal of Physics}\ }\textbf {\bibinfo {volume} {79}},\ \bibinfo
  {pages} {1381} (\bibinfo {year} {2001})}\BibitemShut {NoStop}%
\bibitem [{\citenamefont {Hamilton}\ \emph {et~al.}(2014)\citenamefont
  {Hamilton}, \citenamefont {Lampronti}, \citenamefont {Rowley},\ and\
  \citenamefont {Dutton}}]{Sackville_Hamilton_2014}%
  \BibitemOpen
  \bibfield  {author} {\bibinfo {author} {\bibfnamefont {A.~C.~S.}\
  \bibnamefont {Hamilton}}, \bibinfo {author} {\bibfnamefont {G.~I.}\
  \bibnamefont {Lampronti}}, \bibinfo {author} {\bibfnamefont {S.~E.}\
  \bibnamefont {Rowley}},\ and\ \bibinfo {author} {\bibfnamefont {S.~E.}\
  \bibnamefont {Dutton}},\ }\bibfield  {title} {\bibinfo {title} {Enhancement
  of the magnetocaloric effect driven by changes in the crystal structure of
  \ce{Al}-doped {GGG}, \ce{Gd$_3$Ga${_{5-x}}$Al${_x}$O${_{12}}$}
  (0$\leq$x$\leq$5)},\ }\href {https://doi.org/10.1088/0953-8984/26/11/116001}
  {\bibfield  {journal} {\bibinfo  {journal} {Journal of Physics: Condensed
  Matter}\ }\textbf {\bibinfo {volume} {26}},\ \bibinfo {pages} {116001}
  (\bibinfo {year} {2014})}\BibitemShut {NoStop}%
\bibitem [{\citenamefont {Melot}\ \emph {et~al.}(2009)\citenamefont {Melot},
  \citenamefont {Drewes}, \citenamefont {Seshadri}, \citenamefont
  {Stoudenmire},\ and\ \citenamefont {Ramirez}}]{Melot_SRO_spinels_2009}%
  \BibitemOpen
  \bibfield  {author} {\bibinfo {author} {\bibfnamefont {B.}~\bibnamefont
  {Melot}}, \bibinfo {author} {\bibfnamefont {J.}~\bibnamefont {Drewes}},
  \bibinfo {author} {\bibfnamefont {R.}~\bibnamefont {Seshadri}}, \bibinfo
  {author} {\bibfnamefont {E.}~\bibnamefont {Stoudenmire}},\ and\ \bibinfo
  {author} {\bibfnamefont {A.}~\bibnamefont {Ramirez}},\ }\bibfield  {title}
  {\bibinfo {title} {Magnetic phase evolution in the spinel compounds
  \ce{Zn_{{1-x}}Co_{x}Cr2O4}},\ }\href
  {https://doi.org/10.1088/0953-8984/21/21/216007} {\bibfield  {journal}
  {\bibinfo  {journal} {Journal of Physics: Condensed Matter}\ }\textbf
  {\bibinfo {volume} {21}},\ \bibinfo {pages} {216007} (\bibinfo {year}
  {2009})}\BibitemShut {NoStop}%
\bibitem [{\citenamefont {Karunadasa}\ \emph {et~al.}(2003)\citenamefont
  {Karunadasa}, \citenamefont {Huang}, \citenamefont {Ueland}, \citenamefont
  {Schiffer},\ and\ \citenamefont {Cava}}]{Karunadasa_PNAS}%
  \BibitemOpen
  \bibfield  {author} {\bibinfo {author} {\bibfnamefont {H.}~\bibnamefont
  {Karunadasa}}, \bibinfo {author} {\bibfnamefont {Q.}~\bibnamefont {Huang}},
  \bibinfo {author} {\bibfnamefont {B.~G.}\ \bibnamefont {Ueland}}, \bibinfo
  {author} {\bibfnamefont {P.}~\bibnamefont {Schiffer}},\ and\ \bibinfo
  {author} {\bibfnamefont {R.~J.}\ \bibnamefont {Cava}},\ }\bibfield  {title}
  {\bibinfo {title} {\ce{Ba2LnSbO6} and \ce{Sr2LnSbO6} ({Ln = Dy, Ho, Gd})
  double perovskites: Lanthanides in the geometrically frustrating fcc
  lattice},\ }\href {https://doi.org/10.1073/pnas.0832394100} {\bibfield
  {journal} {\bibinfo  {journal} {Proceedings of the National Academy of
  Sciences}\ }\textbf {\bibinfo {volume} {100}},\ \bibinfo {pages} {8097}
  (\bibinfo {year} {2003})}\BibitemShut {NoStop}%
\bibitem [{\citenamefont {Paddison}\ \emph {et~al.}(2017)\citenamefont
  {Paddison}, \citenamefont {Ehlers}, \citenamefont {Petrenko}, \citenamefont
  {Wildes}, \citenamefont {Gardner},\ and\ \citenamefont
  {Stewart}}]{Paddison_Gd2Sn2O7_2017}%
  \BibitemOpen
  \bibfield  {author} {\bibinfo {author} {\bibfnamefont {J.}~\bibnamefont
  {Paddison}}, \bibinfo {author} {\bibfnamefont {G.}~\bibnamefont {Ehlers}},
  \bibinfo {author} {\bibfnamefont {O.}~\bibnamefont {Petrenko}}, \bibinfo
  {author} {\bibfnamefont {A.}~\bibnamefont {Wildes}}, \bibinfo {author}
  {\bibfnamefont {J.}~\bibnamefont {Gardner}},\ and\ \bibinfo {author}
  {\bibfnamefont {J.}~\bibnamefont {Stewart}},\ }\bibfield  {title} {\bibinfo
  {title} {Spin correlations in the dipolar pyrochlore antiferromagnet
  \ce{Gd2Sn2O7}},\ }\href {https://doi.org/10.1088/1361-648x/aa5d2e} {\bibfield
   {journal} {\bibinfo  {journal} {Journal of Physics: Condensed Matter}\
  }\textbf {\bibinfo {volume} {29}},\ \bibinfo {pages} {144001} (\bibinfo
  {year} {2017})}\BibitemShut {NoStop}%
\bibitem [{\citenamefont {Paddison}\ \emph {et~al.}(2015)\citenamefont
  {Paddison}, \citenamefont {Jacobsen}, \citenamefont {Petrenko}, \citenamefont
  {Fern{\'a}ndez-D{\'\i}az}, \citenamefont {Deen},\ and\ \citenamefont
  {Goodwin}}]{Paddison_GGG_2015}%
  \BibitemOpen
  \bibfield  {author} {\bibinfo {author} {\bibfnamefont {J.}~\bibnamefont
  {Paddison}}, \bibinfo {author} {\bibfnamefont {H.}~\bibnamefont {Jacobsen}},
  \bibinfo {author} {\bibfnamefont {O.}~\bibnamefont {Petrenko}}, \bibinfo
  {author} {\bibfnamefont {M.}~\bibnamefont {Fern{\'a}ndez-D{\'\i}az}},
  \bibinfo {author} {\bibfnamefont {P.}~\bibnamefont {Deen}},\ and\ \bibinfo
  {author} {\bibfnamefont {A.}~\bibnamefont {Goodwin}},\ }\bibfield  {title}
  {\bibinfo {title} {Hidden order in spin-liquid \ce{Gd3Ga5O12}},\ }\href
  {https://doi.org/10.1126/science.aaa5326} {\bibfield  {journal} {\bibinfo
  {journal} {Science}\ }\textbf {\bibinfo {volume} {350}},\ \bibinfo {pages}
  {179} (\bibinfo {year} {2015})}\BibitemShut {NoStop}%
\bibitem [{\citenamefont {Mukherjee}\ \emph {et~al.}(2018)\citenamefont
  {Mukherjee}, \citenamefont {Wu}, \citenamefont {Lampronti},\ and\
  \citenamefont {Dutton}}]{GdBO3_2018}%
  \BibitemOpen
  \bibfield  {author} {\bibinfo {author} {\bibfnamefont {P.}~\bibnamefont
  {Mukherjee}}, \bibinfo {author} {\bibfnamefont {Y.}~\bibnamefont {Wu}},
  \bibinfo {author} {\bibfnamefont {G.~I.}\ \bibnamefont {Lampronti}},\ and\
  \bibinfo {author} {\bibfnamefont {S.~E.}\ \bibnamefont {Dutton}},\ }\bibfield
   {title} {\bibinfo {title} {Magnetic properties of monoclinic lanthanide
  orthoborates, {LnBO${_3}$}, {Ln}={Gd}, {Tb}, {Dy}, {Ho}, {Er}, {Yb}},\ }\href
  {https://doi.org/https://doi.org/10.1016/j.materresbull.2017.10.007}
  {\bibfield  {journal} {\bibinfo  {journal} {Materials Research Bulletin}\
  }\textbf {\bibinfo {volume} {98}},\ \bibinfo {pages} {173} (\bibinfo {year}
  {2018})}\BibitemShut {NoStop}%
\bibitem [{\citenamefont {Ashtar}\ \emph {et~al.}(2021)\citenamefont {Ashtar},
  \citenamefont {Bai}, \citenamefont {Xu}, \citenamefont {Wan}, \citenamefont
  {Wei}, \citenamefont {Liu}, \citenamefont {Marwat},\ and\ \citenamefont
  {Tian}}]{Gd2Be2GeO7_2021}%
  \BibitemOpen
  \bibfield  {author} {\bibinfo {author} {\bibfnamefont {M.}~\bibnamefont
  {Ashtar}}, \bibinfo {author} {\bibfnamefont {Y.}~\bibnamefont {Bai}},
  \bibinfo {author} {\bibfnamefont {L.}~\bibnamefont {Xu}}, \bibinfo {author}
  {\bibfnamefont {Z.}~\bibnamefont {Wan}}, \bibinfo {author} {\bibfnamefont
  {Z.}~\bibnamefont {Wei}}, \bibinfo {author} {\bibfnamefont {Y.}~\bibnamefont
  {Liu}}, \bibinfo {author} {\bibfnamefont {M.}~\bibnamefont {Marwat}},\ and\
  \bibinfo {author} {\bibfnamefont {Z.}~\bibnamefont {Tian}},\ }\bibfield
  {title} {\bibinfo {title} {Structure and {Magnetic} {Properties} of
  {Melilite}-{Type} {Compounds} \ce{RE2Be2GeO7} ({RE} = {Pr}, {Nd},
  {Gd}–{Yb}) with {Rare}-{Earth} {Ions} on {Shastry}–{Sutherland}
  {Lattice}},\ }\href@noop {} {\bibfield  {journal} {\bibinfo  {journal}
  {Inorganic Chemistry}\ }\textbf {\bibinfo {volume} {60}},\ \bibinfo {pages}
  {3626} (\bibinfo {year} {2021})}\BibitemShut {NoStop}%
\bibitem [{\citenamefont {Chen}\ \emph {et~al.}(2014)\citenamefont {Chen},
  \citenamefont {Qin}, \citenamefont {Meng}, \citenamefont {Yang},
  \citenamefont {Wu}, \citenamefont {Fu}, \citenamefont {Zheng}, \citenamefont
  {Liu}, \citenamefont {Tarasenko}, \citenamefont {Orendáč}, \citenamefont
  {Prokleška}, \citenamefont {Sechovský},\ and\ \citenamefont
  {Tong}}]{GdOHCO3_2014}%
  \BibitemOpen
  \bibfield  {author} {\bibinfo {author} {\bibfnamefont {Y.-C.}\ \bibnamefont
  {Chen}}, \bibinfo {author} {\bibfnamefont {L.}~\bibnamefont {Qin}}, \bibinfo
  {author} {\bibfnamefont {Z.-S.}\ \bibnamefont {Meng}}, \bibinfo {author}
  {\bibfnamefont {D.-F.}\ \bibnamefont {Yang}}, \bibinfo {author}
  {\bibfnamefont {C.}~\bibnamefont {Wu}}, \bibinfo {author} {\bibfnamefont
  {Z.}~\bibnamefont {Fu}}, \bibinfo {author} {\bibfnamefont {Y.-Z.}\
  \bibnamefont {Zheng}}, \bibinfo {author} {\bibfnamefont {J.-L.}\ \bibnamefont
  {Liu}}, \bibinfo {author} {\bibfnamefont {R.}~\bibnamefont {Tarasenko}},
  \bibinfo {author} {\bibfnamefont {M.}~\bibnamefont {Orendáč}}, \bibinfo
  {author} {\bibfnamefont {J.}~\bibnamefont {Prokleška}}, \bibinfo {author}
  {\bibfnamefont {V.}~\bibnamefont {Sechovský}},\ and\ \bibinfo {author}
  {\bibfnamefont {M.-L.}\ \bibnamefont {Tong}},\ }\bibfield  {title} {\bibinfo
  {title} {Study of a magnetic-cooling material {Gd}({OH}){CO}$_{3}$},\ }\href
  {https://doi.org/10.1039/C4TA01646G} {\bibfield  {journal} {\bibinfo
  {journal} {J. Mater. Chem. A}\ }\textbf {\bibinfo {volume} {2}},\ \bibinfo
  {pages} {9851} (\bibinfo {year} {2014})}\BibitemShut {NoStop}%
\bibitem [{\citenamefont {Kelly}\ and\ \citenamefont
  {Dutton}(2020)}]{1D_borates_Nicola}%
  \BibitemOpen
  \bibfield  {author} {\bibinfo {author} {\bibfnamefont {N.~D.}\ \bibnamefont
  {Kelly}}\ and\ \bibinfo {author} {\bibfnamefont {S.~E.}\ \bibnamefont
  {Dutton}},\ }\bibfield  {title} {\bibinfo {title} {Magnetic properties of
  quasi-one-dimensional lanthanide calcium oxyborates
  \ce{Ca${_4}$LnO(BO${_3}$)${_3}$}},\ }\href
  {https://doi.org/10.1021/acs.inorgchem.0c01098} {\bibfield  {journal}
  {\bibinfo  {journal} {Inorganic Chemistry}\ }\textbf {\bibinfo {volume}
  {59}},\ \bibinfo {pages} {9188} (\bibinfo {year} {2020})},\ \bibinfo {note}
  {pMID: 32525304}\BibitemShut {NoStop}%
\bibitem [{\citenamefont {Saines}\ and\ \citenamefont
  {Bristowe}(2018)}]{saines_probing_2018}%
  \BibitemOpen
  \bibfield  {author} {\bibinfo {author} {\bibfnamefont {P.~J.}\ \bibnamefont
  {Saines}}\ and\ \bibinfo {author} {\bibfnamefont {N.~C.}\ \bibnamefont
  {Bristowe}},\ }\bibfield  {title} {\bibinfo {title} {Probing magnetic
  interactions in metal–organic frameworks and coordination polymers
  microscopically},\ }\href {https://doi.org/10.1039/C8DT02411A} {\bibfield
  {journal} {\bibinfo  {journal} {Dalton Transactions}\ }\textbf {\bibinfo
  {volume} {47}},\ \bibinfo {pages} {13257} (\bibinfo {year}
  {2018})}\BibitemShut {NoStop}%
\bibitem [{\citenamefont {Hellsvik}\ \emph {et~al.}(2020)\citenamefont
  {Hellsvik}, \citenamefont {Pérez}, \citenamefont {Geilhufe}, \citenamefont
  {Månsson},\ and\ \citenamefont {Balatsky}}]{hellsvik_spin_2020}%
  \BibitemOpen
  \bibfield  {author} {\bibinfo {author} {\bibfnamefont {J.}~\bibnamefont
  {Hellsvik}}, \bibinfo {author} {\bibfnamefont {R.~D.}\ \bibnamefont
  {Pérez}}, \bibinfo {author} {\bibfnamefont {R.~M.}\ \bibnamefont
  {Geilhufe}}, \bibinfo {author} {\bibfnamefont {M.}~\bibnamefont {Månsson}},\
  and\ \bibinfo {author} {\bibfnamefont {A.~V.}\ \bibnamefont {Balatsky}},\
  }\bibfield  {title} {\bibinfo {title} {Spin wave excitations of magnetic
  metal-organic materials},\ }\href
  {https://doi.org/10.1103/PhysRevMaterials.4.024409} {\bibfield  {journal}
  {\bibinfo  {journal} {Physical Review Materials}\ }\textbf {\bibinfo {volume}
  {4}},\ \bibinfo {pages} {024409} (\bibinfo {year} {2020})}\BibitemShut
  {NoStop}%
\bibitem [{\citenamefont {Karunadasa}\ \emph {et~al.}(2005)\citenamefont
  {Karunadasa}, \citenamefont {Huang}, \citenamefont {Ueland}, \citenamefont
  {Lynn}, \citenamefont {Schiffer}, \citenamefont {Regan},\ and\ \citenamefont
  {Cava}}]{Karunadasa_2005}%
  \BibitemOpen
  \bibfield  {author} {\bibinfo {author} {\bibfnamefont {H.}~\bibnamefont
  {Karunadasa}}, \bibinfo {author} {\bibfnamefont {Q.}~\bibnamefont {Huang}},
  \bibinfo {author} {\bibfnamefont {B.~G.}\ \bibnamefont {Ueland}}, \bibinfo
  {author} {\bibfnamefont {J.~W.}\ \bibnamefont {Lynn}}, \bibinfo {author}
  {\bibfnamefont {P.}~\bibnamefont {Schiffer}}, \bibinfo {author}
  {\bibfnamefont {K.~A.}\ \bibnamefont {Regan}},\ and\ \bibinfo {author}
  {\bibfnamefont {R.~J.}\ \bibnamefont {Cava}},\ }\bibfield  {title} {\bibinfo
  {title} {Honeycombs of triangles and magnetic frustration in
  \ce{Sr${L_2}$O$_4$} (${L}$= \ce{Gd}, \ce{Dy}, \ce{Ho}, \ce{Er}, \ce{Tm}, and
  \ce{Yb})},\ }\href {https://doi.org/10.1103/PhysRevB.71.144414} {\bibfield
  {journal} {\bibinfo  {journal} {Phys. Rev. B}\ }\textbf {\bibinfo {volume}
  {71}},\ \bibinfo {pages} {144414} (\bibinfo {year} {2005})}\BibitemShut
  {NoStop}%
\bibitem [{\citenamefont {Numazawa}\ \emph {et~al.}(2003)\citenamefont
  {Numazawa}, \citenamefont {Kamiya}, \citenamefont {Okano},\ and\
  \citenamefont {Matsumoto}}]{DGGvGGG_MCE}%
  \BibitemOpen
  \bibfield  {author} {\bibinfo {author} {\bibfnamefont {T.}~\bibnamefont
  {Numazawa}}, \bibinfo {author} {\bibfnamefont {K.}~\bibnamefont {Kamiya}},
  \bibinfo {author} {\bibfnamefont {T.}~\bibnamefont {Okano}},\ and\ \bibinfo
  {author} {\bibfnamefont {K.}~\bibnamefont {Matsumoto}},\ }\bibfield  {title}
  {\bibinfo {title} {Magnetocaloric effect in
  \ce{(Dy_{{x}}Gd_{{1-x}})3Ga5O_{{12}}} for adiabatic demagnetization
  refrigeration},\ }\href@noop {} {\bibfield  {journal} {\bibinfo  {journal}
  {Physica B: Condensed Matter}\ }\textbf {\bibinfo {volume} {329-333}},\
  \bibinfo {pages} {1656} (\bibinfo {year} {2003})}\BibitemShut {NoStop}%
\bibitem [{\citenamefont {Orendáč}\ \emph {et~al.}(2007)\citenamefont
  {Orendáč}, \citenamefont {Hanko}, \citenamefont {Čižmár}, \citenamefont
  {Orendáčová}, \citenamefont {Shirai},\ and\ \citenamefont
  {Bramwell}}]{Tad_Dy2Ti2O7}%
  \BibitemOpen
  \bibfield  {author} {\bibinfo {author} {\bibfnamefont {M.}~\bibnamefont
  {Orendáč}}, \bibinfo {author} {\bibfnamefont {J.}~\bibnamefont {Hanko}},
  \bibinfo {author} {\bibfnamefont {E.}~\bibnamefont {Čižmár}}, \bibinfo
  {author} {\bibfnamefont {A.}~\bibnamefont {Orendáčová}}, \bibinfo {author}
  {\bibfnamefont {M.}~\bibnamefont {Shirai}},\ and\ \bibinfo {author}
  {\bibfnamefont {S.~T.}\ \bibnamefont {Bramwell}},\ }\bibfield  {title}
  {\bibinfo {title} {Magnetocaloric study of spin relaxation in dipolar spin
  ice {Dy}${_2}${Ti}${_2}${O}${_7}$},\ }\href
  {https://doi.org/10.1103/PhysRevB.75.104425} {\bibfield  {journal} {\bibinfo
  {journal} {Physical Review B}\ }\textbf {\bibinfo {volume} {75}},\ \bibinfo
  {pages} {104425} (\bibinfo {year} {2007})}\BibitemShut {NoStop}%
\bibitem [{\citenamefont {Sosin}\ \emph
  {et~al.}(2005{\natexlab{b}})\citenamefont {Sosin}, \citenamefont {Prozorova},
  \citenamefont {Smirnov}, \citenamefont {Golov}, \citenamefont {Berkutov},
  \citenamefont {Petrenko}, \citenamefont {Balakrishnan},\ and\ \citenamefont
  {Zhitomirsky}}]{Tad_Gd2Ti2O7}%
  \BibitemOpen
  \bibfield  {author} {\bibinfo {author} {\bibfnamefont {S.}~\bibnamefont
  {Sosin}}, \bibinfo {author} {\bibfnamefont {L.}~\bibnamefont {Prozorova}},
  \bibinfo {author} {\bibfnamefont {A.}~\bibnamefont {Smirnov}}, \bibinfo
  {author} {\bibfnamefont {A.}~\bibnamefont {Golov}}, \bibinfo {author}
  {\bibfnamefont {I.}~\bibnamefont {Berkutov}}, \bibinfo {author}
  {\bibfnamefont {O.}~\bibnamefont {Petrenko}}, \bibinfo {author}
  {\bibfnamefont {G.}~\bibnamefont {Balakrishnan}},\ and\ \bibinfo {author}
  {\bibfnamefont {M.}~\bibnamefont {Zhitomirsky}},\ }\bibfield  {title}
  {\bibinfo {title} {Adiabatic demagnetization of a pyrochlore antiferromagnet
  \ce{Gd${_2}$Ti${_2}$O${_7}$}},\ }\href
  {https://doi.org/https://doi.org/10.1016/j.jmmm.2004.11.344} {\bibfield
  {journal} {\bibinfo  {journal} {Journal of Magnetism and Magnetic Materials}\
  }\textbf {\bibinfo {volume} {290-291}},\ \bibinfo {pages} {709} (\bibinfo
  {year} {2005}{\natexlab{b}})},\ \bibinfo {note} {proceedings of the Joint
  European Magnetic Symposia (JEMS' 04)}\BibitemShut {NoStop}%
\bibitem [{\citenamefont {McCusker}\ \emph {et~al.}(1999)\citenamefont
  {McCusker}, \citenamefont {Von~Dreele}, \citenamefont {Cox}, \citenamefont
  {Lou{\"{e}}r},\ and\ \citenamefont {Scardi}}]{McCusker_RietveldRefinement}%
  \BibitemOpen
  \bibfield  {author} {\bibinfo {author} {\bibfnamefont {L.~B.}\ \bibnamefont
  {McCusker}}, \bibinfo {author} {\bibfnamefont {R.~B.}\ \bibnamefont
  {Von~Dreele}}, \bibinfo {author} {\bibfnamefont {D.~E.}\ \bibnamefont {Cox}},
  \bibinfo {author} {\bibfnamefont {D.}~\bibnamefont {Lou{\"{e}}r}},\ and\
  \bibinfo {author} {\bibfnamefont {P.}~\bibnamefont {Scardi}},\ }\bibfield
  {title} {\bibinfo {title} {{Rietveld refinement guidelines}},\ }\href@noop {}
  {\bibfield  {journal} {\bibinfo  {journal} {Journal of Applied
  Crystallography}\ }\textbf {\bibinfo {volume} {32}},\ \bibinfo {pages} {36}
  (\bibinfo {year} {1999})}\BibitemShut {NoStop}%
\bibitem [{\citenamefont {Coelho}(2018)}]{TOPAS_Academic}%
  \BibitemOpen
  \bibfield  {author} {\bibinfo {author} {\bibfnamefont {A.}~\bibnamefont
  {Coelho}},\ }\bibfield  {title} {\bibinfo {title} {{{\it TOPAS} and {\it
  TOPAS-Academic}: an optimization program integrating computer algebra and
  crystallographic objects written in C++}},\ }\href
  {https://doi.org/10.1107/S1600576718000183} {\bibfield  {journal} {\bibinfo
  {journal} {Journal of Applied Crystallography}\ }\textbf {\bibinfo {volume}
  {51}},\ \bibinfo {pages} {210} (\bibinfo {year} {2018})}\BibitemShut
  {NoStop}%
\bibitem [{\citenamefont {Thompson}\ \emph {et~al.}(1987)\citenamefont
  {Thompson}, \citenamefont {Cox},\ and\ \citenamefont
  {Hastings}}]{TCHZ_peaks}%
  \BibitemOpen
  \bibfield  {author} {\bibinfo {author} {\bibfnamefont {P.}~\bibnamefont
  {Thompson}}, \bibinfo {author} {\bibfnamefont {D.~E.}\ \bibnamefont {Cox}},\
  and\ \bibinfo {author} {\bibfnamefont {J.~B.}\ \bibnamefont {Hastings}},\
  }\bibfield  {title} {\bibinfo {title} {{Rietveld refinement of
  Debye{--}Scherrer synchrotron X-ray data from Al${\sb 2}$O${\sb 3}$}},\
  }\href@noop {} {\bibfield  {journal} {\bibinfo  {journal} {Journal of Applied
  Crystallography}\ }\textbf {\bibinfo {volume} {20}},\ \bibinfo {pages} {79}
  (\bibinfo {year} {1987})}\BibitemShut {NoStop}%
\bibitem [{\citenamefont {Kennedy}\ \emph {et~al.}(1997)\citenamefont
  {Kennedy}, \citenamefont {Hunter},\ and\ \citenamefont
  {Howard}}]{kennedy_structural_1997}%
  \BibitemOpen
  \bibfield  {author} {\bibinfo {author} {\bibfnamefont {B.~J.}\ \bibnamefont
  {Kennedy}}, \bibinfo {author} {\bibfnamefont {B.~A.}\ \bibnamefont
  {Hunter}},\ and\ \bibinfo {author} {\bibfnamefont {C.~J.}\ \bibnamefont
  {Howard}},\ }\bibfield  {title} {\bibinfo {title} {Structural and {Bonding}
  {Trends} in {Tin} {Pyrochlore} {Oxides}},\ }\href
  {https://doi.org/10.1006/jssc.1997.7277} {\bibfield  {journal} {\bibinfo
  {journal} {Journal of Solid State Chemistry}\ }\textbf {\bibinfo {volume}
  {130}},\ \bibinfo {pages} {58} (\bibinfo {year} {1997})}\BibitemShut
  {NoStop}%
\bibitem [{\citenamefont {Paddison}\ \emph {et~al.}(2016)\citenamefont
  {Paddison}, \citenamefont {Ong}, \citenamefont {Hamp}, \citenamefont
  {Mukherjee}, \citenamefont {Bai}, \citenamefont {Tucker}, \citenamefont
  {Butch}, \citenamefont {Castelnovo}, \citenamefont {Mourigal},\ and\
  \citenamefont {Dutton}}]{DMSO_monopoles_2016}%
  \BibitemOpen
  \bibfield  {author} {\bibinfo {author} {\bibfnamefont {J.~M.}\ \bibnamefont
  {Paddison}}, \bibinfo {author} {\bibfnamefont {H.}~\bibnamefont {Ong}},
  \bibinfo {author} {\bibfnamefont {J.}~\bibnamefont {Hamp}}, \bibinfo {author}
  {\bibfnamefont {P.}~\bibnamefont {Mukherjee}}, \bibinfo {author}
  {\bibfnamefont {X.}~\bibnamefont {Bai}}, \bibinfo {author} {\bibfnamefont
  {M.}~\bibnamefont {Tucker}}, \bibinfo {author} {\bibfnamefont
  {N.}~\bibnamefont {Butch}}, \bibinfo {author} {\bibfnamefont
  {C.}~\bibnamefont {Castelnovo}}, \bibinfo {author} {\bibfnamefont
  {M.}~\bibnamefont {Mourigal}},\ and\ \bibinfo {author} {\bibfnamefont
  {S.}~\bibnamefont {Dutton}},\ }\bibfield  {title} {\bibinfo {title} {Emergent
  order in the kagome {Ising} magnet \ce{Dy3Mg2Sb3O14}},\ }\href
  {https://doi.org/10.1038/ncomms13842} {\bibfield  {journal} {\bibinfo
  {journal} {Nature Communications}\ }\textbf {\bibinfo {volume} {7}},\
  \bibinfo {pages} {13842} (\bibinfo {year} {2016})}\BibitemShut {NoStop}%
\bibitem [{\citenamefont {Dun}\ \emph {et~al.}(2020)\citenamefont {Dun},
  \citenamefont {Bai}, \citenamefont {Paddison}, \citenamefont {Hollingworth},
  \citenamefont {Butch}, \citenamefont {Cruz}, \citenamefont {Stone},
  \citenamefont {Hong}, \citenamefont {Demmel}, \citenamefont {Mourigal},\ and\
  \citenamefont {Zhou}}]{HMSO_2020_disorder}%
  \BibitemOpen
  \bibfield  {author} {\bibinfo {author} {\bibfnamefont {Z.}~\bibnamefont
  {Dun}}, \bibinfo {author} {\bibfnamefont {X.}~\bibnamefont {Bai}}, \bibinfo
  {author} {\bibfnamefont {J.~A.~M.}\ \bibnamefont {Paddison}}, \bibinfo
  {author} {\bibfnamefont {E.}~\bibnamefont {Hollingworth}}, \bibinfo {author}
  {\bibfnamefont {N.~P.}\ \bibnamefont {Butch}}, \bibinfo {author}
  {\bibfnamefont {C.~D.}\ \bibnamefont {Cruz}}, \bibinfo {author}
  {\bibfnamefont {M.~B.}\ \bibnamefont {Stone}}, \bibinfo {author}
  {\bibfnamefont {T.}~\bibnamefont {Hong}}, \bibinfo {author} {\bibfnamefont
  {F.}~\bibnamefont {Demmel}}, \bibinfo {author} {\bibfnamefont
  {M.}~\bibnamefont {Mourigal}},\ and\ \bibinfo {author} {\bibfnamefont
  {H.}~\bibnamefont {Zhou}},\ }\bibfield  {title} {\bibinfo {title} {Quantum
  versus classical spin fragmentation in dipolar kagome ice
  \ce{Ho3Mg2Sb3O14}},\ }\href {https://doi.org/10.1103/PhysRevX.10.031069}
  {\bibfield  {journal} {\bibinfo  {journal} {Phys. Rev. X}\ }\textbf {\bibinfo
  {volume} {10}},\ \bibinfo {pages} {031069} (\bibinfo {year}
  {2020})}\BibitemShut {NoStop}%
\bibitem [{\citenamefont {Momma}\ and\ \citenamefont
  {Izumi}(2011)}]{momma_vesta_2011}%
  \BibitemOpen
  \bibfield  {author} {\bibinfo {author} {\bibfnamefont {K.}~\bibnamefont
  {Momma}}\ and\ \bibinfo {author} {\bibfnamefont {F.}~\bibnamefont {Izumi}},\
  }\bibfield  {title} {\bibinfo {title} {\textit{{VESTA} 3} for
  three-dimensional visualization of crystal, volumetric and morphology data},\
  }\href {https://doi.org/10.1107/S0021889811038970} {\bibfield  {journal}
  {\bibinfo  {journal} {Journal of Applied Crystallography}\ }\textbf {\bibinfo
  {volume} {44}},\ \bibinfo {pages} {1272} (\bibinfo {year}
  {2011})}\BibitemShut {NoStop}%
\end{thebibliography}%

\end{document}